\documentclass[aps,pre,twocolumn,superscriptaddress,floats,floatfix]{revtex4}
\usepackage[svgnames]{xcolor}
\usepackage{tikz}
\usepackage{pgfplots}
\usepgfplotslibrary{fillbetween}
\usepackage{amssymb,amsmath,amsthm}
\usepackage{graphicx}
\usepackage{epstopdf}
\usepackage{color}
\usepackage{subfigure}
\usepackage{epsfig}
\usepackage{rotating}
\usepackage{mwe}
\usepackage{float}
\usepackage{dcolumn,bm}
\usepackage{verbatim}
\usepackage{hyperref}
\usepackage[normalem]{ulem}

\def\correspondingauthor{\footnote{Author to whom correspondence should be sent: ssanjay@iisc.ac.in }}

\begin{document}
\title{Inertial Particles in Superfluid Turbulence: Coflow and Counterflow}
\author{Sanjay Shukla \correspondingauthor{}}
\affiliation{Centre for Condensed Matter Theory,
	Department of Physics, Indian Institute of Science, Bangalore 560012, India.}
\author{Akhilesh Kumar Verma }
\email{akvermajnusps@gmail.com}
\affiliation{Mathematics Institute, Zeeman Building, University of Warwick, 
	Coventry CV4 7AL, UK.}
\author{Vishwanath Shukla}
\email{vishwanath.shukla@phy.iitkgp.ac.in}
\affiliation{Department of Physics, Indian Institute of Technology Kharagpur,
	Kharagpur 721 302, India.}
\author{Akshay Bhatnagar}
\email{akshayphy@gmail.com}
\affiliation{SeRC (Swedish e-Science Research Centre) and Flow, KTH,\\
	Department of Engineering Mechanics, SE-10044 Stockholm, Sweden.} 
\author{Rahul Pandit }
\email{rahul@iisc.ac.in}
\affiliation{Centre for Condensed Matter Theory, Department of Physics, Indian Institute 
of Science, Bangalore 560012, India.}

\begin{abstract}
We use pseudospectral direct numerical simulations (DNSs) to solve the three-dimensional (3D) Hall-Vinen-Bekharevich-Khalatnikov (HVBK) model of superfluid Helium. We then explore the statistical properties of inertial particles, in both coflow and counterflow superfluid turbulence (ST) in the 3D HVBK system; particle motion is governed by a generalization of the Maxey-Riley-Gatignol equations. We first characterize the anisotropy of counterflow ST by showing that there exist large vortical columns. The light particles show confined motion as they are attracted towards these columns and they form large clusters; by contrast, heavy particles are expelled from these vortical regions. We characterise the statistics of such inertial particles in 3D HVBK ST: (1) The mean angle $\Theta(\tau)$, between particle positions, separated by the time lag $\tau$, exhibits two different scaling regions in (a) dissipation and (b) inertial ranges, for different values of the parameters in our model; in particular, the value of $\Theta(\tau)$, at large $\tau$, depends on the magnitude of ${\bf U}_{ns}$. (2) The irreversibility of 3D HVBK turbulence is quantified by computing the statistics of energy increments for inertial particles. (3) The probability distribution function (PDF) of energy increments is of direct relevance to recent experimental studies of irreversibility in superfluid turbulence; we find, in agreement with these experiments, that, for counterflow ST, the skewness of this PDF is less pronounced than its counterparts for coflow ST or for classical-fluid turbulence.

\end{abstract}

\maketitle
\section{ Introduction}

Over the past few decades, there has been growing interest in studies of the
statistical properties of particles advected by turbulent fluid flows,
especially because of advances in experimental techniques and computational
resources. Such particle advection is of central importance in
geophysical~\cite{Shaw2003, Grabowski2013, Falkovich2002} and
astrophysical~\cite{Armitage2010} flows, industrial process~\cite{Eaton1994,
	Post2002}, nonequilibrium statistical mechanics~\cite{Cardy2008}, and the
visualization of turbulent flows in quantum fluids~\cite{Donnelly1991,
	Paloetti2011, Skrbek2012, Berloff2014, Tsubota2017, BarenghiParker16, Bewley06,
	Bewley08, Poole2005, Mantia2013, Mantia2014, Zmeev2013, Guo14, Yang20}.
However, investigations of particles in turbulent superfluids are in their
infancy, when we compare them with their classical-fluid-turbulence
counterparts. Some experimental groups~\cite{Bewley06, Bewley08, Poole2005,
	Mantia2013, Mantia2014, Zmeev2013, Guo14, Yang20, Mantia2019} have used
particles to visualize vortex lines in superfluid turbulence. In some cases the
particles (e.g., frozen hydrogen or deuterium) are several orders of magnitude
larger than the core size of a vortex. Some of these particles can be modeled
as neutrally buoyant tracer particles in superfluid turbulence. 

Superfluid turbulence is a multiscale problem for which we must use different
levels of description, depending on the length scales that we
consider~\cite{Berloff2014, Tsubota2017, BarenghiParker16}: The
Gross-Pitaevskii equation (GPE)~\cite{Berloff2014, Krstulovic11, Shukla13}
provides a natural description for a low-temperature, weakly-interacting Bose
gas, at length scales comparable to the size of the superfluid vortex core,
which has a healing length $\xi$. The vortex-filament model distinguishes
between individual quantum vortices; but it does not account for the nature of
the vortex core; it is valid on length scales greater than $\xi$, in the
incompressible limit. The Hall-Vinen-Bekharevich-Khalatnikov (HVBK) two-fluid
model does not resolve individual quantum vortices, but uses macroscopic,
classical-vorticity fields (this assumes local polarization of quantum-vortex
lines). At the level of kinetic theory, there is the model of Zaremba, Nikuni, and Griffin~\cite{ZNG99}. Some groups have begun to
investigate the interactions of classical particles with vortices in a GPE
description of superfluid turbulence~\cite{Winiecki00, Shukla16, Shukla18,
	Giuriato19, GPK19, GKN19}. These particles are active in the sense that they
affect the superfluid flow while they are advected by this flow.

Within the HVBK framework, we can consider both coflow and counterflow
Superfluid Turbulence (ST). In coflow ST, the two fluids move in the same
direction, with the same mean velocities; in counterflow ST, superfluid and
normal-fluid components move in opposite directions because of an imposed
temperature gradient. The statistical properties of counterflow ST are
different from those of classical-fluid turbulence~\cite{sahoo} and coflow
ST~\cite{Uns,Polanco20,Lvov21}. In counterflow-ST experiments, there is a
steady flow along a channel that is closed at one end and open at the other
end; a heat flux $q$ is generated by passing a current through a resistor, at
the closed end; the normal fluid, which carries this heat flux, with velocity
${\bf{U}}_n = q/\rho ST$, entropy per unit mass $S$, and at temperature $T$,
moves away from the closed end; to maintain zero mass flux, $ \rho_s {\bf U}_s
+ \rho_n {\bf U}_n = 0 $, so the superfluid flows in the opposite direction,
towards the closed end, with velocity ${\bf{U}}_s = -(\rho_n/\rho_s) {\bf{U}}_n$. Thus, there is a relative
velocity ${\bf{U}}_{ns} = {\bf{U}}_{n}-{\bf{U}}_{s} $ between the two fluids in
such thermally driven counterflow ST\cite{Carlo}
\begin{eqnarray}
{\bf{U}}_{ns} = \frac{\rho}{\rho_s} {\bf{U}}_{n}
\end{eqnarray}

where $\rho_n$, $\rho_s$ are the densities of the normal-fluid and superfluid component respectively and $\rho = \rho_n  + \rho_s$ is the total density. So long as the heat flux $q$
is small, this counterflow is laminar; but if $q$ increases beyond a critical
value, this flow is turbulent.

We carry out a systematic study
of inertial particles in 3D HVBK coflow ST and counterflow
ST~\cite{Donnelly1999, Barenghi1983, Hall1956, Khalatnikov1965}.  This model
has been studied, without particles, for superfluid $^4$He in both two
dimensions (2D) and 3D~\cite{Roche2009, Shukla2015, Biferale2019, Verma2019};
moreover, a recent study~\cite{Giorgio2020} has investigated the clustering of
inertial particles in 3D HVBK turbulence. Below the critical temperature $T_{c}
= 2.17$K, superfluid $^4$He is thought of as comprising a viscous, normal-fluid
component and an inviscid superfluid component. The density ratio of the
normal-fluid component and the total fluid $(\rho_n/\rho)$ is equal to one at
$T_c$; and $\rho_n/\rho$ decreases as we decrease the temperature $T$; at $T=0$
the normal-fluid component vanishes and $^4$He is completely in the superfluid
form. For $0 < T < T_c$, the normal-fluid component interacts with quantized
vortices of the superfluid, via mutual friction~\cite{Vinen1957}; and this
causes dissipation in the superfluid component. In this study, we consider
inertial particles, whose size is smaller than the Kolmogorov dissipation
length of the normal fluid; we assume that these particles are passive insofar as they do not
affect the flow and their turbulence-induced accelerations are much larger
than the acceleration because of gravity. 

We use pseudospectral direct numerical simulations (DNSs)  to solve a
simplified  version of the 3D HVBK model, with inertial particles, whose
statistical properties we then study for different values of the
mutual-friction coefficients, in this model, and for various values of the
Stokes numbers ($St = \tau_{p}/ \tau_{f}$, where, $\tau_{p}$ is the
particle-response time and $\tau_{f}$ the Kolmogorov-dissipation time scale of
the fluid). Inertial particles are different from Lagrangian tracer particles,
which follow the fluid velocity; because of their inertia, particles
cluster for $St \simeq 1 $ \cite{Toschi2005, Biferale2005, Giorgio2020}. We summarize our principal results before we present the details of our work:
 
\begin{enumerate}

\item We characterize the anisotropy of counterflow ST by using
spectra~\cite{Lvov21} and the anisotropy tensor (see below); we then
calculate particle statistics by employing the measures given below.

\item We define persistence times, based on the velocity-gradient tensors of
the normal fluid and the superfluid, and show that the cumulative probability distribution functions (CPDFs) of these
persistence times have exponential tails in different regions of the flow.

\item The mean angle $\Theta(\tau)$, between particle positions separated by
the time lag $\tau$ (defined precisely below), has two different
scaling regions (in dissipation and inertial ranges) for
different values of the Stokes numbers and the mutual-friction coefficients. 

\item The CPDFs of the curvature $\kappa$ and the modulus $\theta$ of the
torsion of particle trajectories have power-law tails with universal
exponents, which are independent of all the control parameters in our model.

\item We characterize the irreversibility of 3D HVBK turbulence, by using
inertial particles, and quantify its dependence on the Stokes numbers. 

\end{enumerate}

The remainder of this paper is organised as follows. We describe the HVBK model
and our DNSs in Sec.~\ref{Sec:Model}. We present, in Sec.~\ref{Sec:Result},
the details of our results.
 We discuss the implications of our results in the concluding Sec.~\ref{Sec:Conclusions}.

\section{ Model and Numerical Simulations}
\label{Sec:Model}
We use the simplified form of the 3D HVBK equations~\cite{Roche2009}. In
addition to the kinematic viscosity $\nu_{\rm n}$ of the normal fluid, we
include Vinen's effective viscosity~\cite{vinen2002quantum} $\nu_{\rm s}$ in
the superfluid component to mimic the dissipation because of (a) vortex
reconnections and (b) interactions between superfluid vortices and the normal
fluid~\cite{2007bottleneck}; The equations for this simplified, incompressible 3D HVBK model (we use the form suggested in
Ref.~\cite{Uns}) for fluctuations ${\bf u}_n$ and ${\bf u}_s$ with zero mean are:

\begin{eqnarray}
{\partial_{t}{\bf u}_{n}} + [({\bf u}_{n} +{\bf U}_{n}) \cdot \nabla] {\bf u}_{n} &=& -\frac{1}{\rho_n} \nabla p_n + \nu_n \nabla^2 {\bf u}_{n} \nonumber  \\ &+& 
{\bf F}^{n}_{mf} + {\bf f}_{n} , \nonumber \\               
{\partial_{t} {\bf  u}_{s}} + [({\bf u}_{s} +{\bf U}_{s}) \cdot \nabla] {\bf u}_{s} &=& -\frac{1}{\rho_s} \nabla p_s + \nu_s \nabla^2 {\bf u}_{s} \nonumber \\ &+& {\bf F}^{s}_{mf} + {\bf f}_{s}  . 
\label{eq:HVBK2}
\end{eqnarray}
Here, ${\bf u}_{n} ({\bf u}_{s})$, ${\bf U}_{n} ({\bf U}_{s})$, $\rho_{n}
(\rho_{s})$, $p_{n} (p_{s})$ and  $\nu_{n}(\nu_{s})$ are the velocity, mean
velocity, density, pressure, and kinematic viscosity of the normal fluid
(superfluid), respectively; ${\bf U}_{n}$ and ${\bf U}_{s}$ vanish for 
\textit{coflow} but not for \textit{counterflow}. The mean relative velocity ${\bf U}_{ns}={\bf U}_n-{\bf U}_s$ is non-zero for \textit{counterflow} and it cannot be eliminated by a
Galilean transformation as discussed in Ref.~\cite{Polanco20}.
The mutual-friction terms $ {\bf F}^{n}_{mf} $ and ${ \bf F}^{s}_{mf} $, which
lead to energy transfer between normal-fluid and superfluid
components~\cite{Morris2008,Wacks2011}, are
\begin{eqnarray}
{\bf F}^{s}_{mf} &=& -\frac{\rho_n}{\rho} {\bf f}_{mf}; \;
{\bf F}^{n}_{mf} = \frac{\rho_s}{\rho} { \bf f}_{mf}; \nonumber \\
{\bf f}_{mf} &=& \frac{B}{2}\widehat{\omega}_s \times({\bf \omega}_{s}
\times ({\bf u}_{n} - {\bf u}_{s} )) \nonumber \\
&+& \frac{B'}{2}{\bf \omega}_s
\times ( {\bf u}_{n}-{\bf u}_{s}) ,
\label{eq:MFterms}
\end{eqnarray}
where $ \rho = \rho_{n} + \rho_{s} $ is the total density, ${\bf u}_{ns} = {\bf u}_{n} - {\bf u}_{s} $ the slip velocity, $ \omega_{s} =
\nabla \times {\bf u}_{s} $ the superfluid vorticity, $B$ and $B'$ the
mutual-friction coefficients, and $ {\bf f}_{n}$ and $ {\bf f}_{s}$ the
external forcing terms for the normal fluid and superfluid, respectively,
and the caret denotes a unit vector. We consider incompressible flows for 
which we use the incompressibility conditions
\begin{eqnarray}
\nabla \cdot {\bf u}_{n} &=& 0 \,\,\,\, {\rm and} \nonumber \\
\nabla\cdot {\bf u}_{s} &=& 0 ,
\label{eq:incomp}
\end{eqnarray}

for the normal fluid and the superfluid, respectively. Given these incompressibility conditions, the pressures 
$p_n$ and $p_s$ can be eliminated from the equations; if these pressures
are required, we can calculate them by using the Poisson equations that relate them to the velocity fields, but we do not need them in this study.
We carry out a Fourier-pseudospectral DNS study of the 3D HVBK
equations~\ref{eq:HVBK2} and \ref{eq:incomp} by using the following:

\begin{itemize}
	\item a cubical box of side $2\pi$, with periodic boundary conditions along
	each direction, $N^{3}$ collocation points, and the $2/3$ dealiasing
	rule~\cite{Canuto1998}. 
	
	\item in this pseudospectral
		method~\cite{Krstulovic11,Orszag}, the derivatives in
		Eq.~\ref{eq:HVBK2} are evaluated in Fourier space where they
		are local, and products are evaluated in physical space; for
		Fast Fourier transforms (FFT) and their inverses we use the
		FFTW~\cite{FFTW} libraries:
	
	\item the constant-energy-injection scheme~\cite{Lamorgese2005, Sahoo2011} is
	used to force the Fourier modes, which lie in the first two shells in
	Fourier space, for both the normal fluid and the superfluid; 
	
	\item the second-order Adams-Bashforth scheme for time
	marching~\cite{Sahoo2011}.  
	
	\item In our direct numerical simulations (DNSs), we use
smooth initial conditions; furthermore, the flow is incompressible, so there
are no shocks. Of course, we do use $2/3$ dealiasing, as we have mentioned in
our paper; we have checked explicitly, by using two resolutions, namely,
$N^3=256^3$ and $N^3=512^3$ that the statistical properties we consider  are
not affected significantly by this change of resolution.
	
\end{itemize}

The parameters for our DNSs are given in Table~\ref{tableI}; here, $Re_\lambda^n (Re_\lambda^s)$, 
	$\tau_{eddy}^n (\tau_{eddy}^s)$, $\eta_{n} (\eta_{s})$, 
	$\tau^n_{\eta} (\tau^s_{\eta})$, and $T$ are the Taylor-microscale Reynolds number, eddy-turn-over time, Kolmogorov 
	dissipation length and time scales for the normal fluid (superfluid), and 
	temperature (in Kelvin), respectively. We use the
temperature-dependent values of $\rho_{n}, \rho_{s}, B $, and $B'$ from the
experiments of Ref.~\cite{Donnelly1998}. The values of the viscosities are taken from Ref.~\cite{viscosity}. We use $\nu_n/\nu_s \leq 10$; it
is difficult to go beyond this ratio with the resolution of
our DNS. [This is similar to the problem faced by DNSs
of magnetohydrodynamics (MHD) turbulence when the
magnetic Prandtl number (the ratio of the fluid kine-
matic viscosity and magnetic diffusivity ) is very different from unity~\cite{Sahoo2011}.]

To study the advection of inertial particles in this HVBK model, we
consider that (a) the radius of the particles $a \ll \eta_{n}$,
where $\eta_{n}$ is the Kolmogorov dissipation length scales
for normal fluid, (b) particles do not interact with
each other, 
(c) particles do not affect the fluid flows, and (d)
turbulence-induced particle accelerations are much greater than the
acceleration because of gravity. The particle's radius $\simeq 1 \to 5 \times 10^{-3}cm$;
and the Kolmogorov length scale for normal fluid at $T=1.65K$ is $\simeq
1\times 10^{-2}cm$. Under these conditions, the evolution equations for the particles, discussed in Refs.~\cite{Gatignol1983, Maxey1983, Bec2006} for a classical fluid, can be generalized, in the HVBK
model~\cite{Poole2005}, to:

\begin{eqnarray}
\frac{d {\bf v}(t)}{dt} &=& \frac{ {\bf u}_{n}({\bf X},t) - {\bf v}(t)}{\tau_{p}} 
+ \beta\left(\frac{\rho_n}{\rho} \frac{D{\bf u}_n}{Dt} + \frac{\rho_s}{\rho} \frac{D{\bf u}_s}{Dt}  \right) \nonumber \\
\frac{d {\bf X}(t)}{dt} &=& {\bf v}(t) ;
\label{eq:MRG}
\end{eqnarray}
here, ${\bf v}(t)$ and ${\bf X}(t)$  are, respectively, the velocity and
position of the particle at time $t$; and ${\bf u}_{n}({\bf X},t)$ and ${\bf u}_{s} ({\bf X},t)$ are the Eulerian normal-fluid and superfluid velocities at
position $\bf X$ and time $t$; $D/Dt$ is the material derivative; the term with the coefficient
\begin{equation}
\beta \equiv 3\rho/(2\rho_p + \rho)
\label{eq:beta}
\end{equation}
accounts for added-mass effects ($\rho_p$ is the particle's density); the
particle-response time for the normal fluid is:
\begin{eqnarray}
\tau_{p}=\frac{a^{2}\rho} {3\beta \rho_n\nu_{n}}; 
\label{eq:tau}
\end{eqnarray}

To study the statistical properties of such particles, we solve
Eq.~\ref{eq:MRG} for (a) $N_p = 100000$ particles, by using the first-order
Euler scheme for time marching and tri-linear interpolation, to calculate the
particles' velocities at off-grid points, and (b) for different Stokes numbers
\begin{eqnarray}
{\rm{St_n}} = \frac{\tau_{p}} {\tau_{n}} 
\end{eqnarray}
with $\tau_{n} =(\nu_{n}/\epsilon_{n})^{\frac{1}{2}}$ the Kolmogorov time scale for the normal fluid and $\epsilon_n$ is the rate of kinetic energy dissipation for the normal fluid; the higher these Stokes
numbers, the higher the particle inertia.

\begin{table*}[!hbt]
\resizebox{1.0\linewidth}{!}
{
\begin{tabular}{|l|l|l|l|l|l|l|l|l|l|l|l|l|l|l|l|l|l|l|l|}
\hline
$Run$ & $N$ & $T$ & $\rho_n/\rho $ & $\widetilde{U}_{ns}$& $B$ & $ B'$ & $\nu_n/10^{4} $ & $\nu_s/10^{4}$ & $dt/10^{4}$ & $f_{n}$& $f_{s}$  &  $ Re_\lambda^n $ & $  Re_\lambda^s $ &$ \tau_{eddy}^n  $ & $  \tau_{eddy}^s $ & $ k_{max} \eta_n $ & $ k_{max} \eta_s$ & $\tau^n_{\eta}$ & $ \tau^s_{\eta}$ \\
\hline
{$\bf R1 $} & $256$ &$1.65$& $0.193$ &0.00 &$1.14$ & $0.15$ & $11.3$ &
$2.3$ &$10$ & $ 0.02 $ & $ 0.02 $ &  $34 $ &$124$ & $1.43$ &$1.34$ & $1.26 $ & $0.46$ & $ 0.19 $ & $ 0.13 $ \\ 
\hline
{$\bf R2 $} & $256$ &$2.10$& $0.741$ & 0.00&$1.30$ & $-0.07$ & $1.67$ &
$10$ &$10$ & $ 0.02 $ & $ 0.02 $ &  $153$ &    $40$ & $1.30$ &$1.41$ & $0.36$ & $1.14$ & $ 0.11 $ & $ 0.18 $ \\ 
\hline
{$\bf R3 $} & $512$ & $1.65$&$0.193$ &0.00 &$1.14$ & $0.15$ & $11.3$ &
$2.3$ &$8$ & $ 0.02 $ & $ 0.02 $ &  $35 $ &    $211$ & $1.24$ &$1.14$ & $2.04 $ & $0.49$ & $ 0.15 $ & $ 0.09 $ \\ 
\hline
{$\bf R4 $} &$256$ &$1.65$& $0.193$& $11.32$ &$1.14$  & $0.15$ &$11.3 $ & $2.3 $ &$10  $  & $0.02$ & $0.02$ & $287$ & $758$ & $1.60$ & $1.45$ & $2.45$ & $0.68$  & $0.58$ & $0.29$ \\
{$\bf R5 $}& & & &$20.31$  & & & & & & & &$237$ &$561$ & $1.90$ & $1.65$ & $2.46$ & $0.71$ & $0.74$ & $0.31$\\
\hline
{$\bf R6 $} & $256$ &$2.10$& $0.741$ & $8.94$&$1.30$ & $-0.07$ & $1.67 $ & $10 $ & $10$ & $0.02$ & $0.02$& $374$ & $177$ & $1.82$ & $2.10$ &$0.50$  & $2.14$ & $0.21$ & $0.63$  \\
\hline
{$\bf R7 $} & $ 512 $ & $1.65$& $ 0.193 $ & $12.90$&$1.14$ & $0.15$ & $11.3 $ & $2.3 $ & $8  $ & $0.02$ & $0.02$& $58$  & $244$ & $4.32$ & $2.51$ & $6.22$ & $1.77$ & $1.17$ & $0.47$   \\
\hline
{$\bf R8 $} & $512$ &$2.10$& $0.741$ & $8.54$&$1.30$ & $-0.07$ & $1.7 $ & $10.0 $ & $8$ &$ 0.02$ & $0.02$& $190$ & $62 $& $2.17$ &$ 4.12$ &$1.13$  & $5.40$ & $0.26$ & $1.00$  \\
\hline
\end{tabular} }
\caption{\label{tableI} \small Parameters for our DNS runs. Coflow ST: 
	$\bf R1 $ - $\bf R3 $; and counterflow ST: $\bf R4 $ - $\bf R8 $.
	$N^3$ is the total no of collocation points; $\rho_{n}/\rho$ is the 
	normal-fluid fraction; the non-dimensionalized counterflow velocity 
	$\widetilde{U}_{ns} = |{\bf U}_{ns}|/u_T^n$, where  
	$u_T^n = \sqrt{\langle|{\bf u}_n|^2\rangle}$ and the angular brackets denote
	the average over the turbulent, but statistically steady, state of the 
	3D HVBK system; $B$ and $B'$ are the coefficients of mutual friction; 
	$\nu_{n} (\nu_{s})$, $Re_\lambda^n (Re_\lambda^s)$, 
	$\tau_{eddy}^n (\tau_{eddy}^s)$, $\eta_{n} (\eta_{s})$, 
	$\tau^n_{\eta} (\tau^s_{\eta})$, and $T$ are the kinematic
	viscosity, Taylor-microscale Reynolds number, eddy-turn-over time, Kolmogorov 
	dissipation length and time scales for the normal fluid (superfluid), and 
	temperature (in Kelvin), respectively; the time step is $dt$; $k_{max}$ is the 
	largest wave number (after dealiasing); and $f_{n}$ ($f_{s}$) provide constant 
	energy injection into the first two  shells in Fourier space for the normal 
	fluid (superfluid); we force both the fluids. }
\end{table*}

\section{Results}
\label{Sec:Result}

\begin{figure*}[!hbt]
\color{blue}\boxed{
	\includegraphics[width=0.25\linewidth]{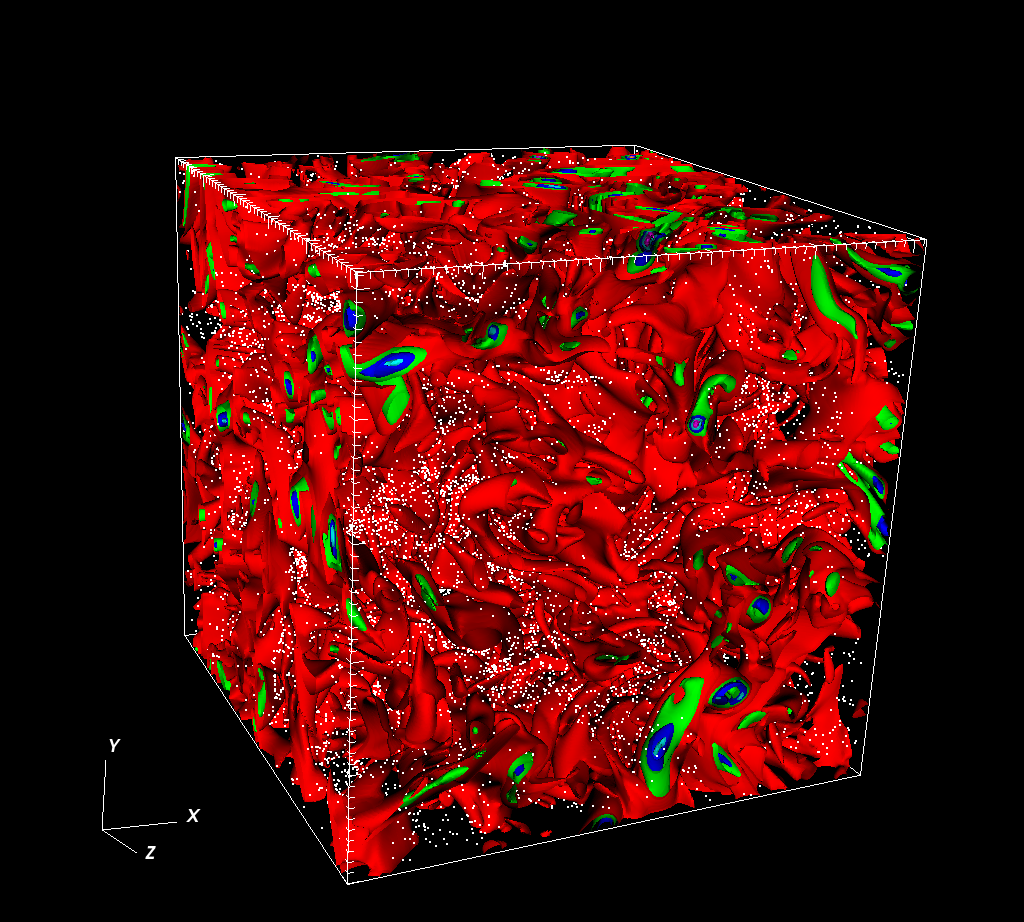}}\put(-80,120){Coflow}
\put(-110,100){\color{white}(a)}
\color{red}\boxed{
	\includegraphics[width=0.25\linewidth]{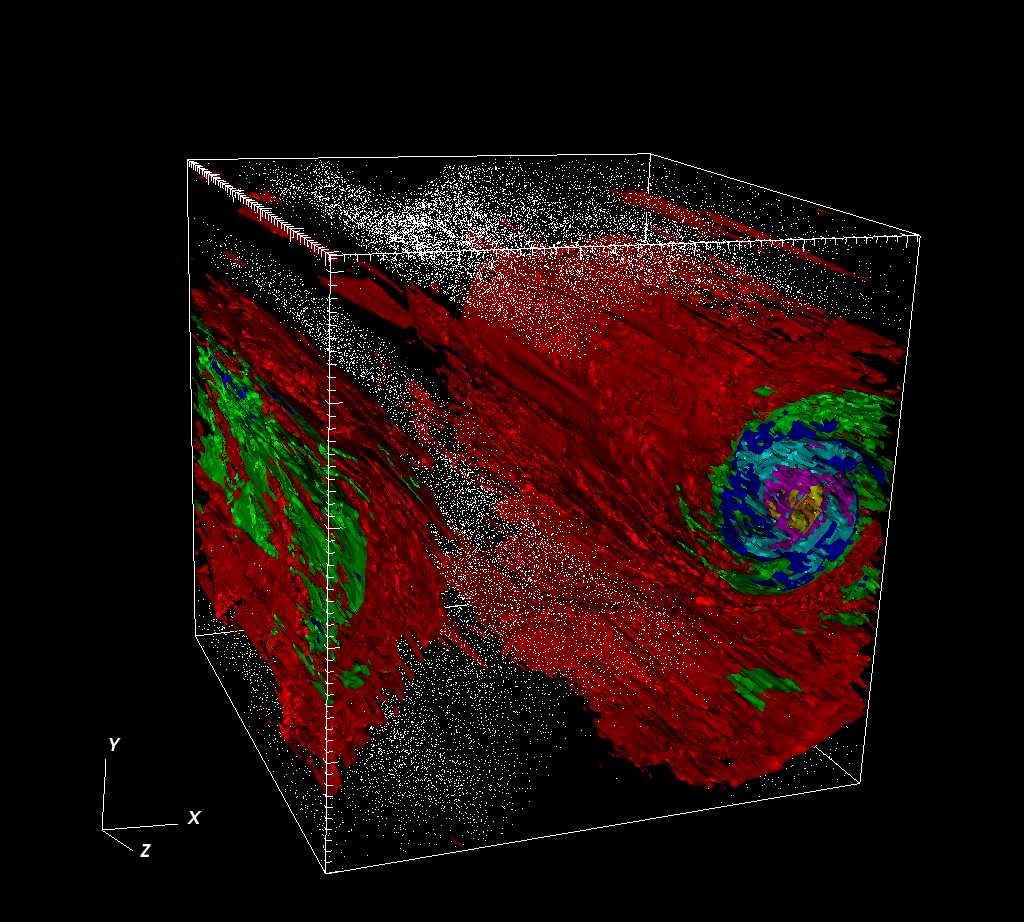}
	\includegraphics[width=0.25\linewidth]{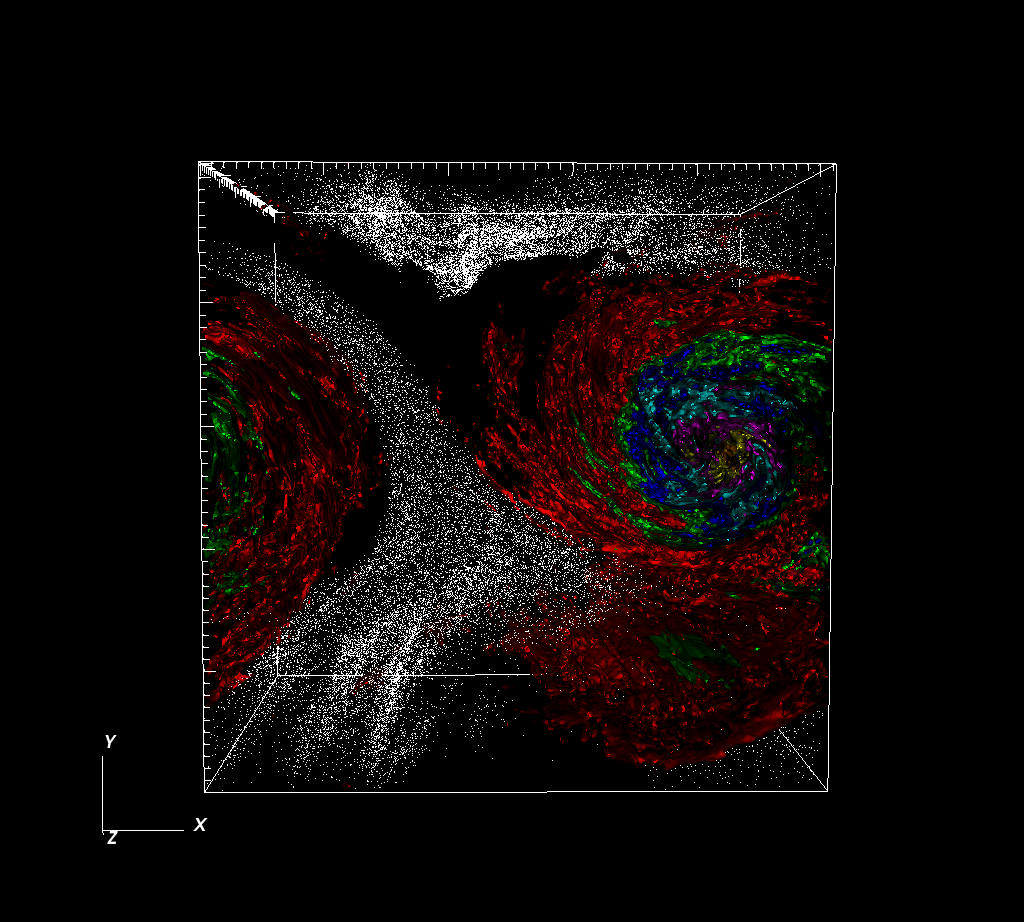}
     \includegraphics[width=0.25\linewidth]{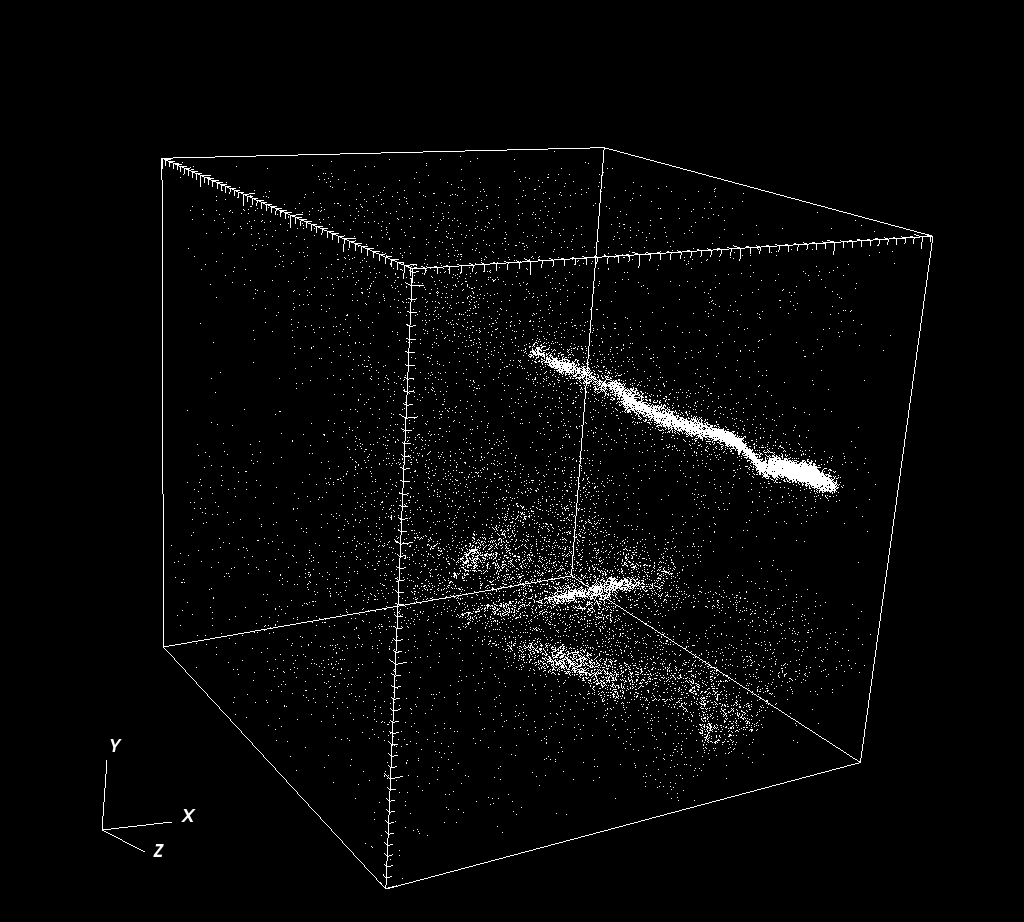}}\put(-220,120){Counterflow}
     \put(-365,100){\color{white}(b)}
     \put(-238,100){\color{white}(c)}
     \put(-115,100){\color{white}(d)}

    \color{black}
	\caption{Isosurface plots of the magnitude of the normal-fluid vorticity
		$|\omega_n|$ with particles as white points, at $T=1.65$K. {\bf (a)} coflow ST({\bf R1}) and {\bf (b)}-{\bf (d)} counterflow ST ({\bf R5}); {\bf(c)} a 2D view of {\bf (b)} for heavy particles ($\beta=0.1$); and {\bf (d)} the same plot as in {\bf(c)} but for light particles ($\beta=1.25$). The counterflow velocity is along the $z$-direction, which points out of the page in {\bf (c)}.}
	\label{fig:iso_surface}
\end{figure*}

We study the statistics of inertial particles for different values of ${\rm St}_{\rm n}$ and $\beta$ in the 3D HVBK model for our different DNS
runs. Before we discuss these statistics of inertial particles we present, in
Fig.~\ref{fig:iso_surface}, isosurface plots of the magnitude of the normal-fluid
vorticity $|\omega_n|$ at $T=1.65$K: Figures~\ref{fig:iso_surface}(a) and (b)-(d) show, respectively, such isosurface plots for coflow and counterflow ST; in the latter case, the counterflow velocity points along $\widehat{\bf U}_{ns} = \hat{e}_k$, where $\hat{e}_k$ is the unit vector along the $z$ direction. We present isosurfaces  for $\rm St_n=1.0$ and $\beta = 1.25 \, (\rho_{p}/\rho=0.7)$ and $\beta = 0.1 \, (\rho_{p}/\rho=14.5)$. For coflow, the spatial 
organization of isosurfaces appears to be isotropic at $T=1.65K$ and particles form clusters [Fig.~\ref{fig:iso_surface}(a)] as in classical fluid turbulence. In contrast, counterflow ST
exhibits large-scale vortex columns (Fig.\ref{fig:iso_surface}(b)), in which 
heavy particles ($\beta=0.1$) form large clusters [Fig.~\ref{fig:iso_surface}(c)] that 
are repelled from the regions with large vortical structures; however, light particles 
($\beta=1.25$) are attracted towards these structures [Fig.~\ref{fig:iso_surface}(d)]. In Fig.\ref{fig:sup_iso} of the Supplementary Material\ref{Sec:supplementary} , we show isosurface plots of $|\omega_n|$
for counterflow ST at $T=2.10K$; the distribution of particles is similar to that at $T=1.65K$.

We also characterize the anisotropy of counterflow ST by using the anisotropy 
tensor $a_{ij}$ and energy spectra. The anisotropy tensor has the components
\begin{equation} a_{ij} = \frac{\overline{u_i u_j}}{\overline{u_iu_i}} -
\frac{1}{3}\delta_{ij},
\label{eq:anistensor}
\end{equation}
where $u_i$ and $u_j$ are the Cartesian components of the fluctuating velocity
for the normal fluid, we use the Einstein summation convention for repeated indices, and
the overbar denotes the volume average. We calculate different off-diagonal components of
$a_{ij}$ and find, e.g., that $a_{ij} \simeq 10^{-3}$ for coflow ST at $T =
1.65K$; by contrast, for counterflow ST, $a_{ij} \simeq 2\times 10^{-1}$.  This
shows clearly the degree of anisotropy in the counterflow ST in our DNS.
Furthermore, we examine the anisotropy of counterflow ST by using the following
energy spectra~\cite{Lvov21}:

\begin{eqnarray}
E^l_{k_{\parallel}} &=& \frac{1}{2} \sum_{k_{\parallel}-\frac{1}{2} < k'<
	k_{\parallel}+\frac{1}{2}} \tilde{{\bf u}}^l_{\parallel}({\bf k}')
\cdot \tilde{{\bf u}}^l_{\parallel}(-{\bf k}'); \nonumber \\
E^l_{k_{\perp}} &=& \frac{1}{2} \sum_{k_{\perp}-\frac{1}{2} < k'<
	k_{\perp}+\frac{1}{2}} \tilde{{\bf u}}^l_{\perp}({\bf k}') \cdot
\tilde{{\bf u}}^l_{\perp}(-{\bf k}');
\label{eq:spectra}
\end{eqnarray}
here, $l$ can be $n$ or $s$; we denote by $\tilde{{\bf u}}^l_{\parallel}$ and
$\tilde{{\bf u}}^l_{\perp}$ the spatial Fourier transforms of the velocities in
the directions ${\bf k}_{\parallel}$ and ${\bf k}_{\perp}$, respectively, where
${\bf k}_{\parallel} = ({\bf k} \cdot {\bf \widehat{U}}_{ns}) {\bf \widehat{U}}_{ns}$
and, perpendicular to it, ${\bf k}_{\perp} = {\bf k} - {\bf k}_{\parallel}$;
and $k', \, k_{\parallel}$, and $k_{\perp}$ are, respectively, the magnitudes
of ${\bf k}'$, ${\bf k}_{\parallel}$, and ${\bf k}_{\perp}$. 
\begin{figure}[!hbt]
	\includegraphics[width=1\linewidth]{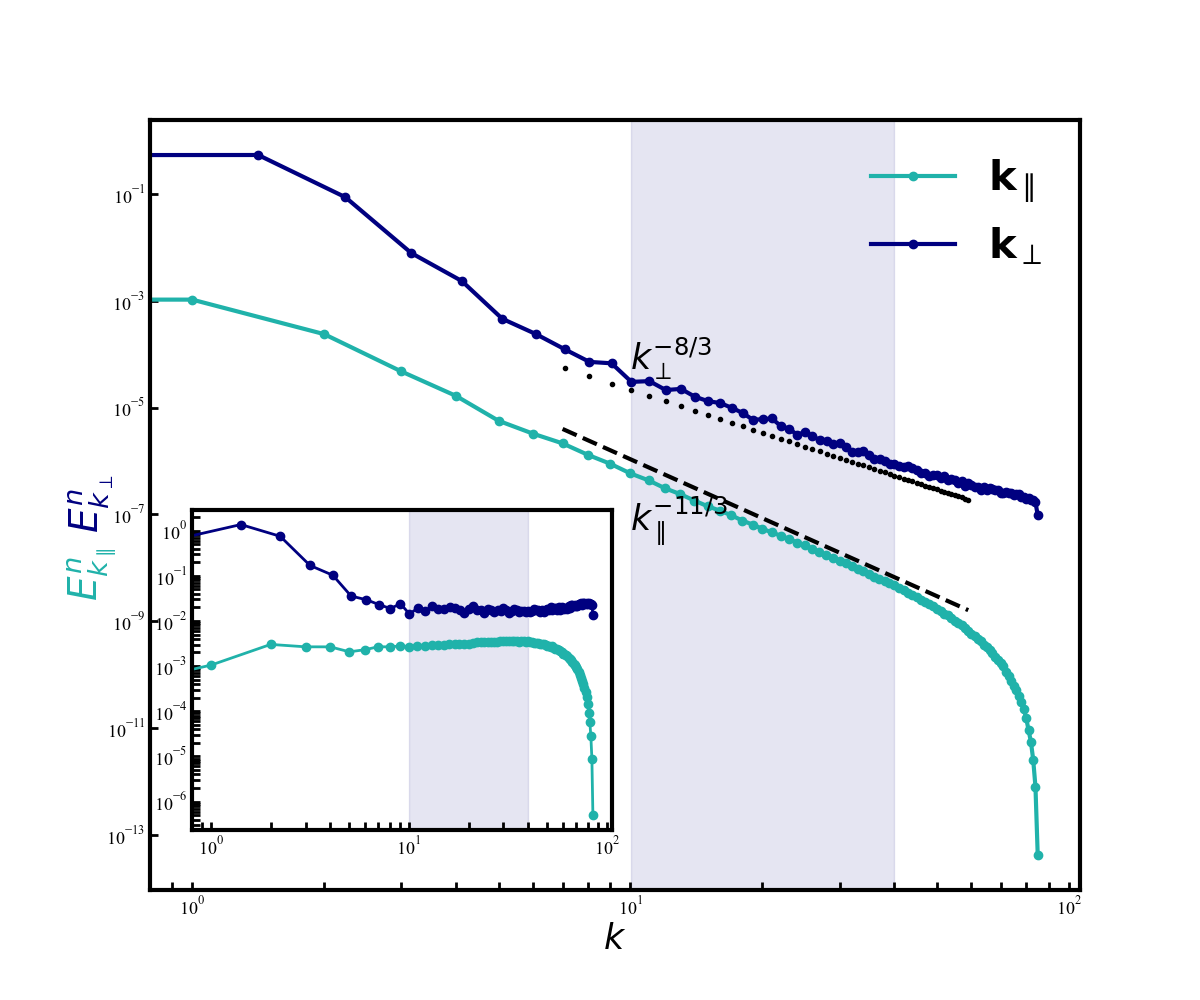}
	\caption{Log-log plots of the energy spectra (Eq.~\ref{eq:spectra})
		$E^n_{k_{\perp}}$ (dark blue) and $E^n_{k_{\parallel}}
		$ (cyan) for  the normal fluid component of counterflow ST with $T = 1.65 K$ and $\widehat{\bf U}_{ns} = \widehat{e_k}$ (run $\bf R5$). The blue shaded region shows the inertial
		range of scales. In the inset, we plot the compensated spectra $E^n_{k_{\perp}}*k_{\perp}^{8/3}$ (dark blue) and $E^n_{k_{\parallel}}
		* k_{\parallel}^{11/3}$ (cyan).
	}	
	\label{fig:aniso_spectra}
\end{figure}
We plot, in Fig.\ref{fig:aniso_spectra}, the compensated energy
spectra $E^n_{k_{\perp}}*k_{\perp}^{8/3}$ (dark blue) and $E^n_{k_{\parallel}}
* k_{\parallel}^{11/3}$ (cyan) for the normal-fluid component of counterflow ST at $T = 1.65 K$ for ${\bf
	\widehat{U}}_{ns} =  \widehat{e_k}$
(run $\bf R5$). Note that $E^n_{k_{\parallel}}$ is strongly suppressed relative to
$E^n_{k_{\perp}}$; furthermore, these spectra show two distinct
(blue-shaded region) power-law forms that are consistent with $E^n_{k_{\perp}}
\sim k_{\perp}^{-8/3}$ and $E^n_{k_{\parallel}}\sim k_{\parallel}^{-11/3}$
. These spectra are in agreement with the recent results of Ref.~\cite{Lvov21}.

This anisotropy of counterflow ST affects the trajectories of inertial particles, which are advected by such turbulence. We can visualise this qualitatively by including the positions of, say, $10000$ particles (shown via small white spheres) along with the isosurfaces, in Fig.\ref{fig:iso_surface}, of the magnitude of the normal-fluid vorticity $|\omega_n|$.  Clearly, in the case of counterflow ST at $T=1.65$K ((c) of Fig.~\ref{fig:iso_surface}), particles form large clusters around large vortical structures and 
move principally along the direction $\widehat{\bf U}_{ns}$ of the counterflow velocity.

In Subsection~\ref{subsec:QR} we characterize the flow in the Eulerian frame by using 
joint PDFs (JPDFs) of the $Q$ and $R$ invariants of the velocity-gradient tensor. In Subsection~\ref{subsec:ID_Increments} we obtain the angle $\Theta$ that quantifies the statistics of inertial-particle displacement increments. Subsection~\ref{subsec:Trajectories} is
devoted to a characteriation of the statistical properties of the geometry of
particle trajectories. In Subsection~\ref{subsec:Irreversibility}, we
characterize the irreversibility of 3D HVBK turbulence. In all these
Subsections we compare and contrast our results for coflow ST and counterflow
ST; we also examine the dependence of some of the results on the non-dimensionalized counterflow velocity $\widetilde{U}_{ns} = |{\bf U}_{ns}|/u_T^n$, where $u_T^n = \sqrt{\langle|{\bf u}_n|^2\rangle}$ and the angular brackets denote the average over the turbulent, but statistically steady, state of the 3D HVBK system. Figure~\ref{fig:steady_kinetic} in the Appendix shows the time series of the  volume-averaged energy, $E(t)= \sum_{k_{\parallel}}E_{k_{\parallel}} + \sum_{k_{\perp}}E_{k_{\perp}}$, in the statistically steady state for run {\bf R1}; here $E_{k_{\parallel}}$ and $E_{k_{\perp}}$ are defined in Eqs.\ref{eq:spectra}. 

\subsection{Joint probability distribution of Q-R invariants}
\label{subsec:QR}
We begin by calculating the invariants $P_{\rm a}$, $Q_{\rm a}$, and $R_{\rm
	a}$ of the velocity-gradient tensor $\mathcal {A}_{{\rm a}{ij}}={\partial_j
	u_{{\rm a}i}} $:
\begin{eqnarray}
P_{\rm a} &=& -{\rm Tr}(\mathcal {A}_{\rm a }); \nonumber \\
Q_{\rm a} &=& -\frac {1}{2}{\rm Tr }(\mathcal {A}_{\rm a}^2);  \nonumber \\
R_{\rm a} &=& -\frac{1}{3}{\rm Tr }(\mathcal {A}_{\rm a}^3); 
\label{eq:PQR}
\end{eqnarray}

where the subscript a stands for $n$ or $s$, and $i,j = 1, 2, 3$.  For
incompressible flows, $P_{\rm a} = 0$.  The discriminant for the characteristic
equation of $\mathcal  {A}_{\rm a}$ is 

\begin{equation}
\Delta_{\rm a} = \frac{27}{4}R_{\rm a}^2 + Q_{\rm a}^3.
\end{equation}

We use these invariants and $\Delta_{\rm a }$, in the $Q_{\rm a}-R_{\rm a}$
plane, to characterize the following four types of flows regions (for this 
well-established method see, e.g., Ref.~\cite{Akshaypersistence}
and references therein): 

\begin{itemize}
	\label{subsec:Persistence}
	\item {\bf Region A}: vortical flow with stretching, for $\Delta_{\rm a} > 0$ and $R_{\rm a} < 0$; 
	\item {\bf Region B}: vortical flow with compression, for $\Delta_{\rm a} > 0$ and $R_{\rm a} > 0$; 
	\item {\bf Region C}: flow with biaxial strain, for $\Delta_{\rm a} < 0$ and $R_{\rm a} < 0$; 
	\item and {\bf Region D}: flow with axial strain, for $\Delta_{\rm a} < 0$ and $R_{\rm a} > 0$. 
\end{itemize}

\begin{figure*}
	\color{blue}\boxed{
	\includegraphics[width=0.24\linewidth]{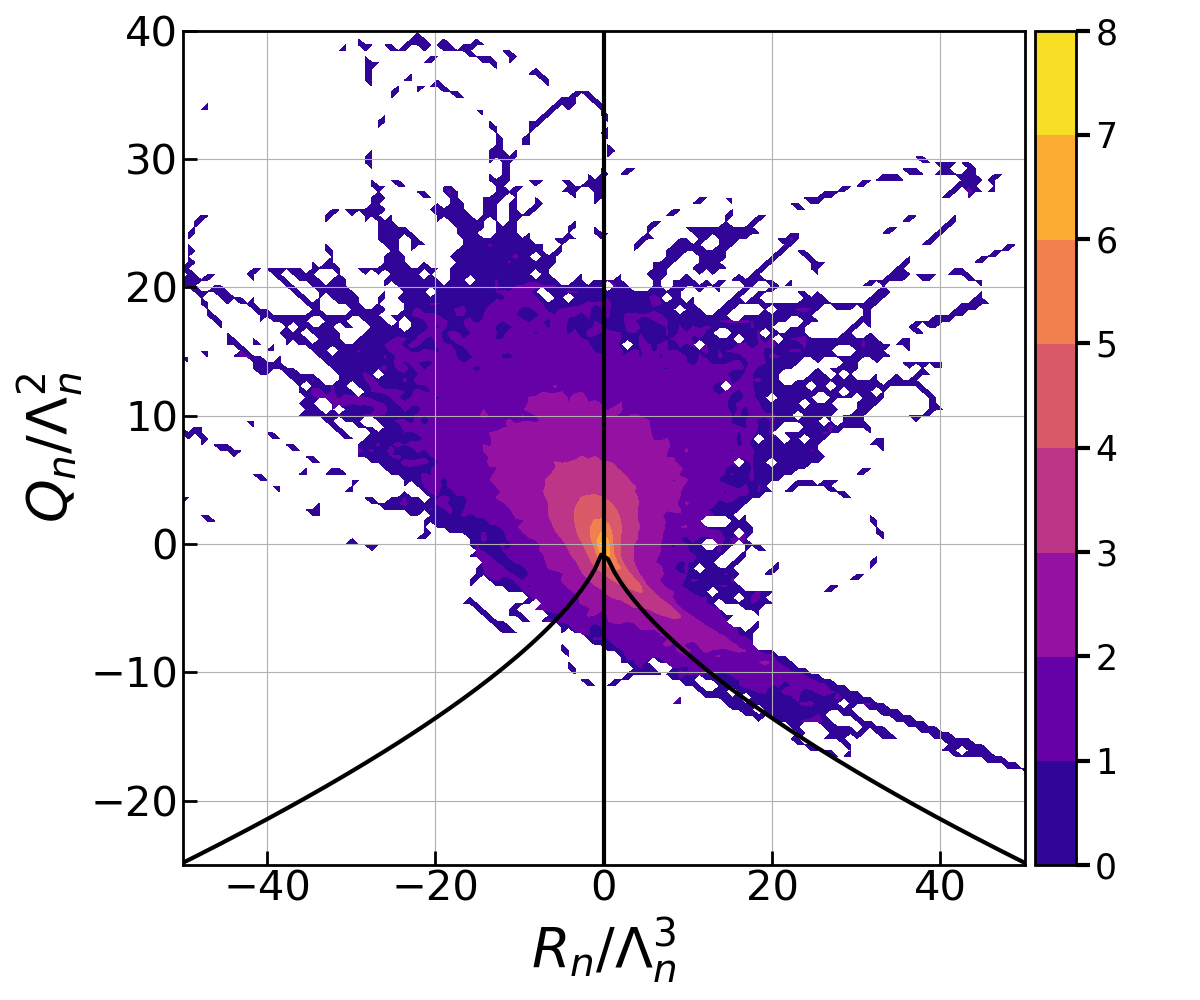} \put(-100,90){\color{black} (a)}
	\put(-40,40){\color{black} \bf(B)}
	\put(-100,40){\color{black} \bf(A)}
	\put(-80,20){\color{black} \bf(C)}
	\put(-60,20){\color{black} \bf(D)}
	\includegraphics[width=0.24\linewidth]{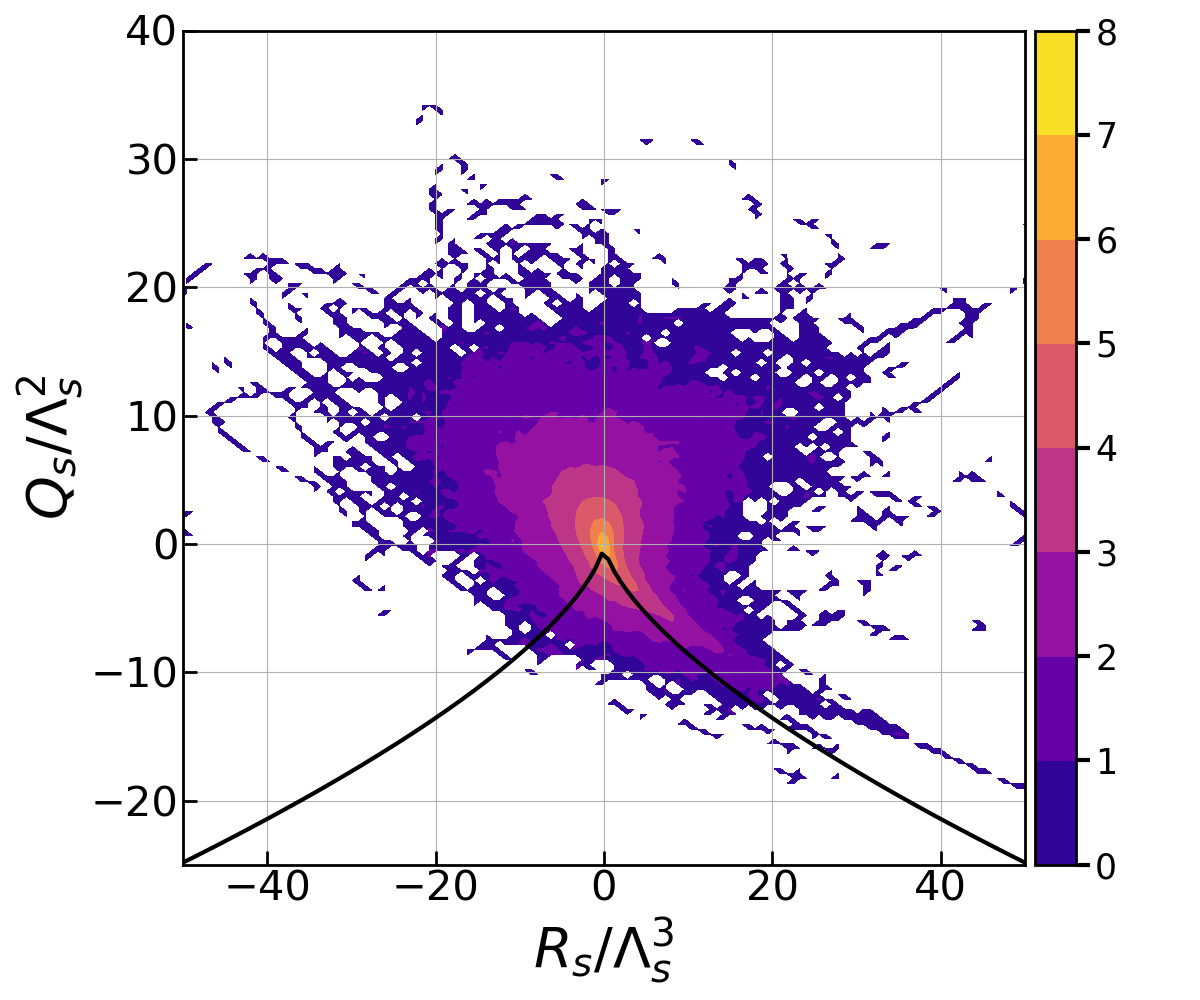}
    \put(-100,90){ \color{black} (b)}}\put(-140,110){\bf  Coflow} 
   \color{red}\boxed{
	\includegraphics[width=0.24\linewidth]{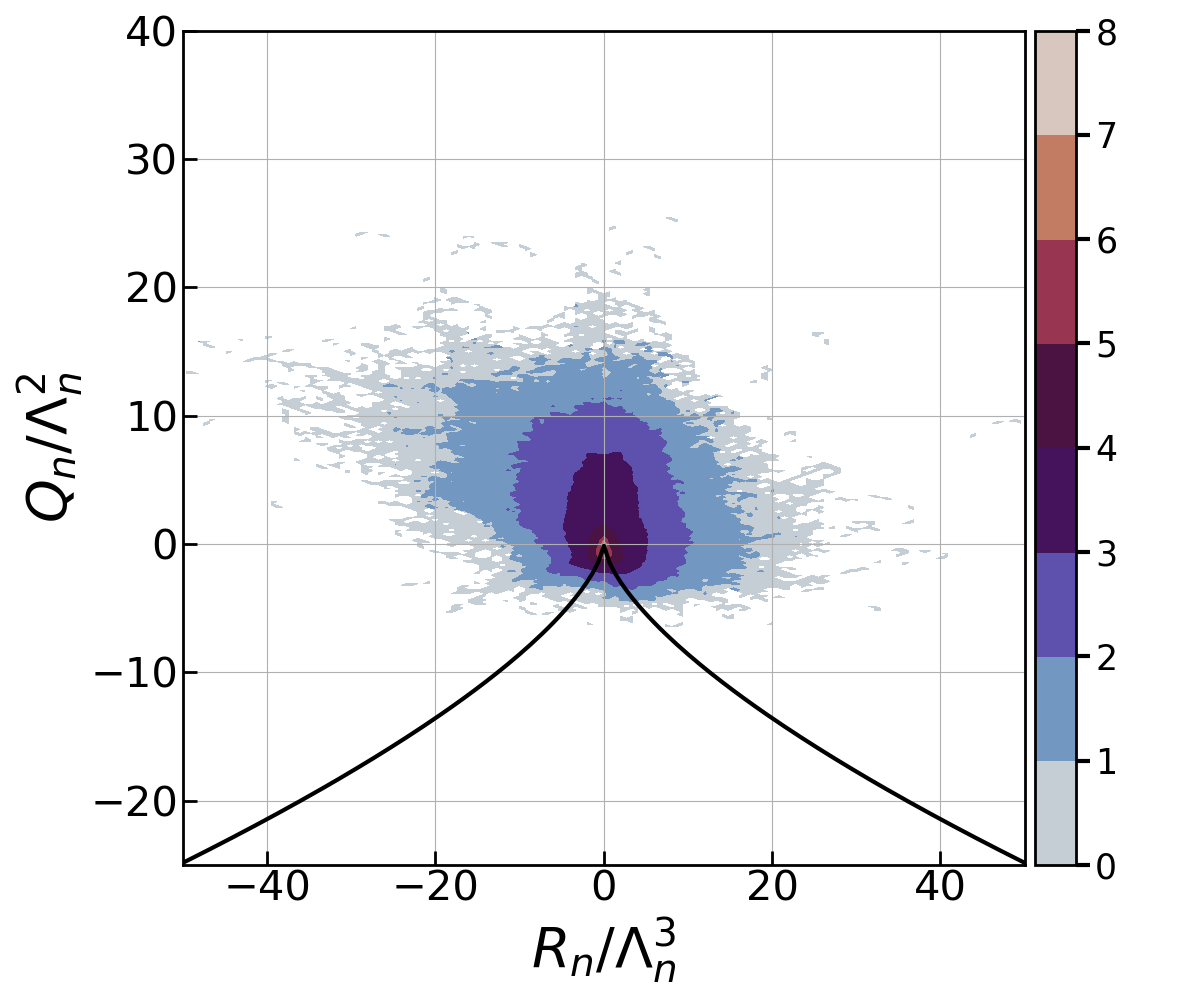} 
	\put(-100,90){\color{black} (c)}
	\includegraphics[width=0.24\linewidth]{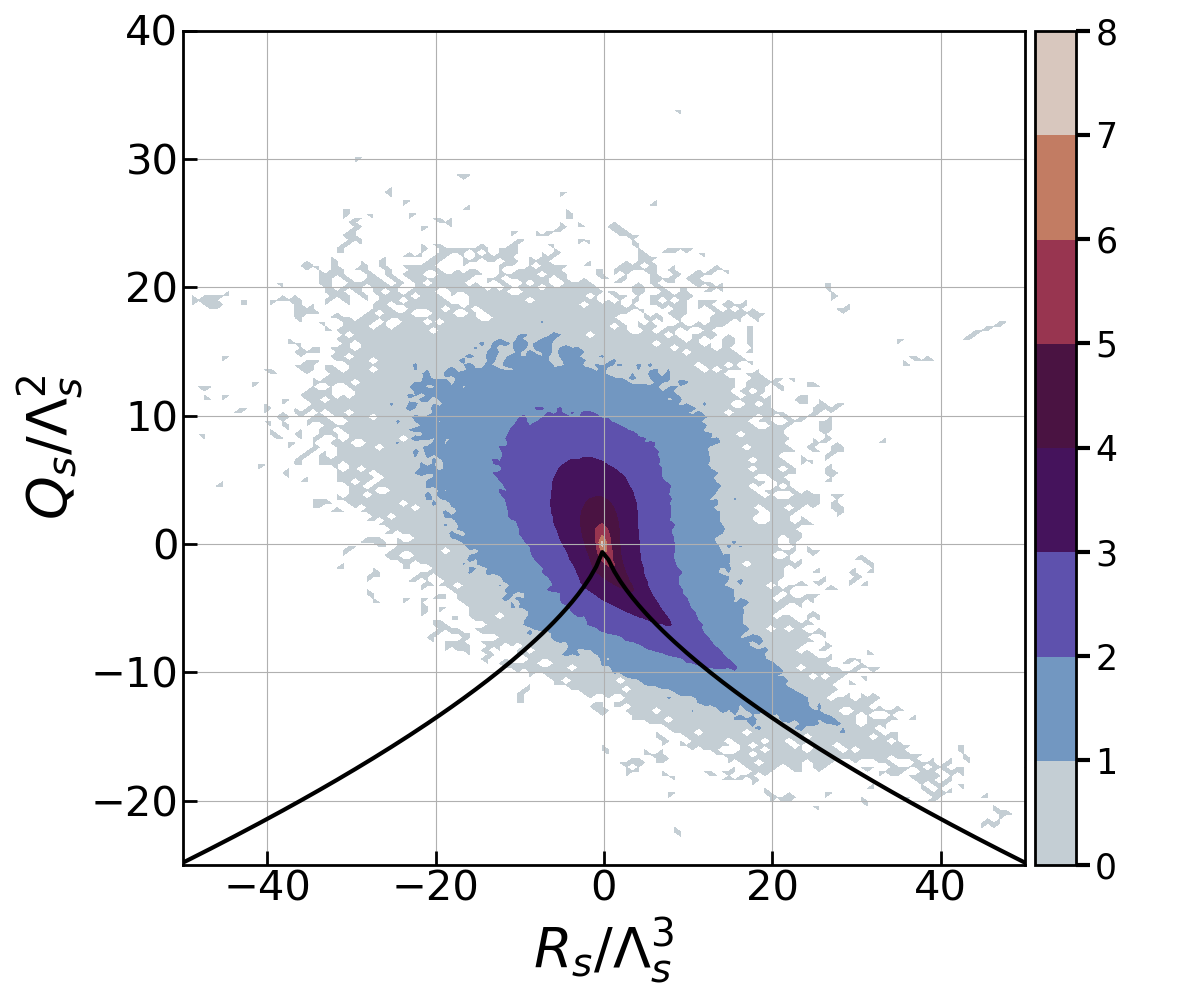}
    \put(-100,90){ \color{black} (d)}}\put(-145,110){\bf  Counterflow} 
\color{black}
	\caption {Filled contour plots of four representative
		JPDFs for coflow ST  [(a) and (b) from run {\bf{R3}}] and counterflow ST
		[(c) and (d) from run {\bf{R7}}]; we give JPDFs of $Q_{\rm n}$ and
		$R_{\rm n}$ in the first and third column and of $Q_{\rm s}$ and $R_{\rm s}$ in 
		the second and fourth column; these are in the Eulerian frame.
		$\Lambda_n = \frac{u_{\eta_n}}{\eta_n}$ and
		$\Lambda_s = \frac{u_{\eta_s}}{\eta_s}$.}
	\label{fig:jpdfs}
\end{figure*}

Joint PDFs (JPDFs) of $Q_a$ and $R_a$ are often used to characterize turbulent
flows in classical-fluid turbulence~\cite{Akshaypersistence}, where they have a
characteristic \textit{tear-drop} shape, i.e., in
strain-dominated regions ($Q<0$), $R>0$ is more probable than $R<0$, whereas
the opposite holds in vortical regions ($Q>0$). In Fig.~\ref{fig:jpdfs} we
present filled contour plots of four representative JPDFs for coflow ST
(Fig.~\ref{fig:jpdfs} (a) and (b)) and counterflow ST (Fig.~\ref{fig:jpdfs} (c)
and (d)) at $T=1.65 K$; these are in the Eulerian frame. The four flow regions, (A)-(D), are shown in Fig.\ref{fig:jpdfs}(a).
We note that the JPDFs for coflow ST have a tear-drop shape, as in
classical-fluid turbulence; but those for counterflow ST show some deviations
from this shape,  which means that, in the strain-dominated	region ($Q<0$), both $R<0$ and $R>0$ are almost equally probable (and likewise
for the vortical region ($Q>0$)). Some groups~\cite{velocity_gradient} have
found, for various experimental turbulent flows, that the shape of the $Q-R$
JPDF depends on the flow and that deviations from a tear-drop shape may arise
if we have vortex-sheet-like structures rather than vortex-tube-like
structures; these depend on the sign of the second eigen value of strain-rate
tensor. We will discuss this in detail, in the context of counterflow ST,
elsewhere. In this paper,  we focus principally on our particle-based studies.

In each one of these flow regions, (A)-(D), we calculate the PDFs of
persistence times $t^{per}_n$ and $t^{per}_s$ for the normal-fluid ($n$) and
superfluid ($s$) components, respectively. These are the times spent by a
particle, in a given region, before it moves to another region. [For
classical-fluid turbulence, see Ref.~\cite{Akshaypersistence}]. We calculate
persistence-time PDFs in the Eulerian frame, by measurements of $Q_{\rm a},
R_{\rm a}$ and $\Delta_{\rm a}$, at a fixed point in space, as a function of
time $t$. We get similar PDFs for tracers or inertial particles by
following the trajectory of each such particle and obtaining  $Q_{\rm a},
R_{\rm a}$ and $\Delta_{\rm a}$ along its trajectory.

\begin{figure}
	\includegraphics[width=0.45\linewidth]{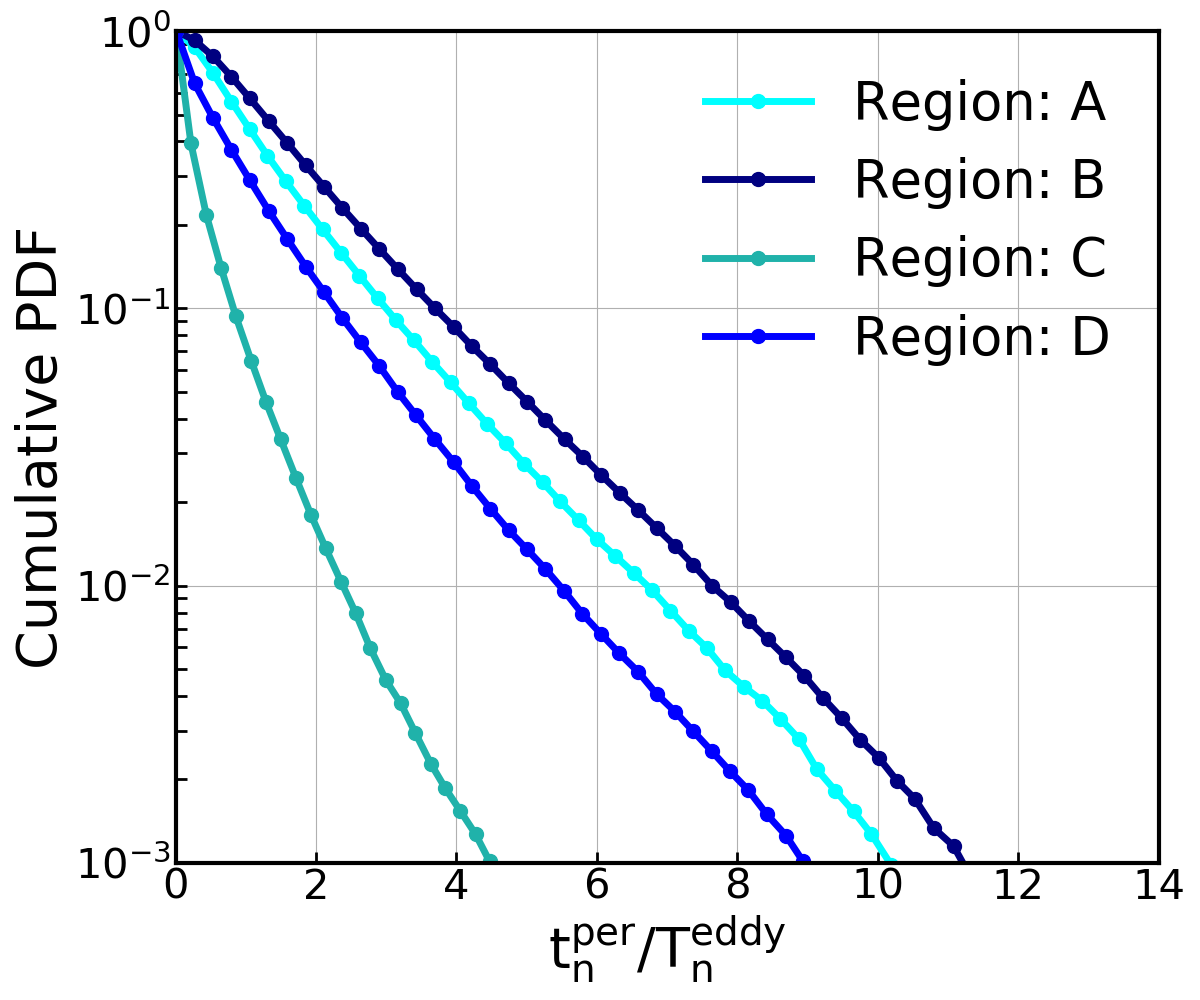}
	\put(-80,80){ (a)}
	\put(-65,90){ Coflow}
	\includegraphics[width=0.45\linewidth]{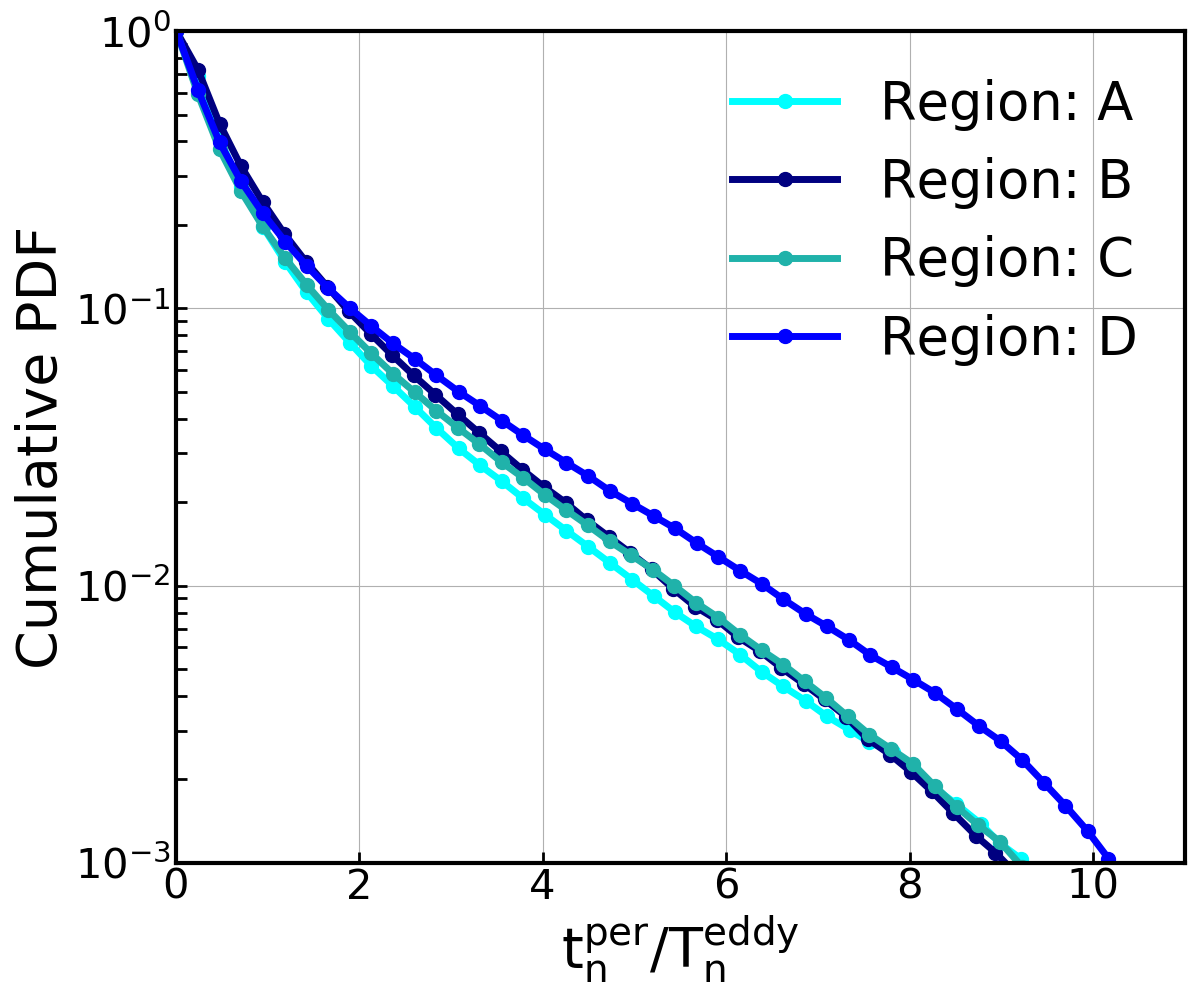}
	\put(-80,80){ (b)}
	\put(-75,90){ Counterflow}
	\caption{Semilog plots of the CPDFs of the persistence times, $t^{per}_n$, at $T=1.65K$ in the  Eulerian frame for the normal-fluid ($n$) component; for coflow ST (run {\bf R3}) in (a) and for counterflow ST (run {\bf R7}) in (b).}
	\label{fig:eul_per}
\end{figure}

In Fig.~\ref{fig:eul_per} we present semilog plots of the persistence-time CPDFs at two temperatures ($T=1.65$K and $2.10$K), in the Eulerian frame, for the normal fluid  and for coflow ST in Fig.~\ref{fig:eul_per}(a) and for counterflow ST in Fig.\ref{fig:eul_per}(b). We give similar plots for the superfluid component, in Fig.\ref{fig:sup_eul_per}, in the Supplementary Material\ref{Sec:supplementary}. From the semilog plots in Figs.~\ref{fig:eul_per} and \ref{fig:sup_eul_per}, we observe that, for  both coflow and counterflow ST, persistence-time CPDFs (and PDFs) have exponentially decaying tails in all the regions A-D and in both the normal fluid and the superfluid.

\subsection{Inertial-particle Displacement Increments}
\label{subsec:ID_Increments}
 In the context of
	classical-fluid turbulence, it has been noted in Ref.~\cite{Bos2015} that the
	study of the changes in direction  of Lagrangian tracers reveals two power-law
	ranges. We carry out the analog of this analysis for inertial particles
	advected by 3D HVBK coflow and counterflow ST; our study highlights the effect
	of $\widetilde{U}_{ns}$ on the change in direction of these particles. From
our DNSs, we obtain the angle $\Theta(\tau)$ between subsequent
inertial-particle-displacement increments~\cite{Bos2015} as a function of the
time lag $\tau$ as follows:
\begin{equation}
\delta {\bf X(x}_{0},t,\tau) = {\bf X(x}_{0},t)- 
{\bf X(x}_{0},t-\tau),
\end{equation}
where ${\bf X(x}_{0},t)$ is the position of the particle at time $t$, 
$ {\bf x}_{0}$ is the reference position for the particle at time 
$t_{0}$. The angle $\Theta(t,\tau)$ is given by 
\begin{equation}
\cos(\Theta(t,\tau)) = \frac{\delta{\bf X(x}_0,t,\tau)\cdot
	{\delta\bf X(x}_0,t+\tau,\tau)}{{\left|\delta {\bf X(x}_{0},t,
		\tau)\right|}{\left|\delta{\bf X(x}_{0},t+\tau,\tau)\right |}},
\label{eq:Theta}
\end{equation}
whose average value, over the time $t$ and the number of particles $N_p$, is 
\begin{equation}
\Theta(\tau) = \langle {\left|\Theta(t,\tau) \right|} \rangle_{t,N_P}. 
\label{eq:ThetaAv}
\end{equation}

\begin{figure*}[!hbt]
	\color{blue}\boxed{\color{black}
	\includegraphics[width=0.3\linewidth]{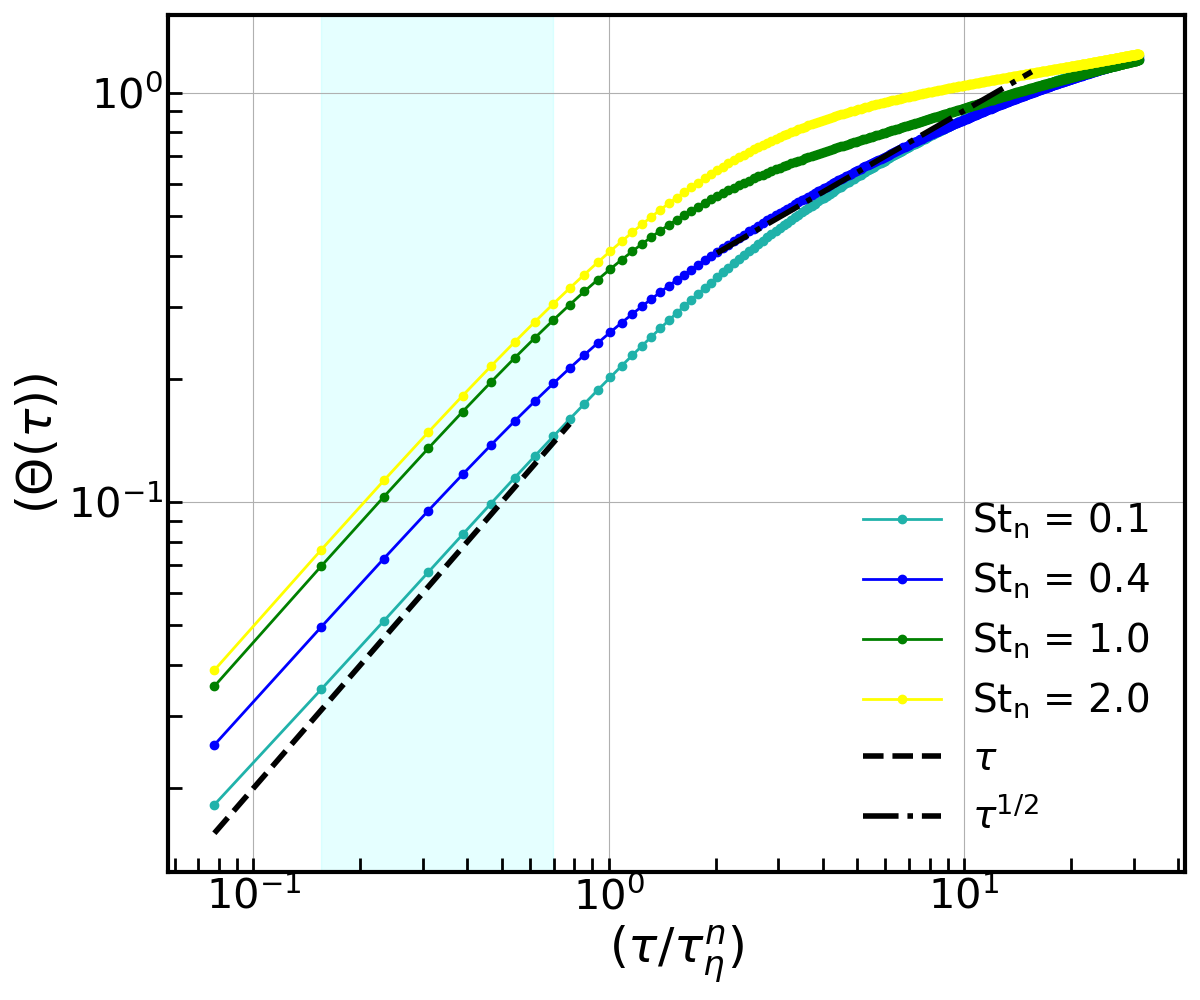}
		\put(-120,105){\rm {\bf(a)}: T = 1.65 K}
	    	}\put(-100,140){\bf  Coflow}
	\color{red}\boxed{\color{black}	
	\includegraphics[width= 0.3\linewidth]{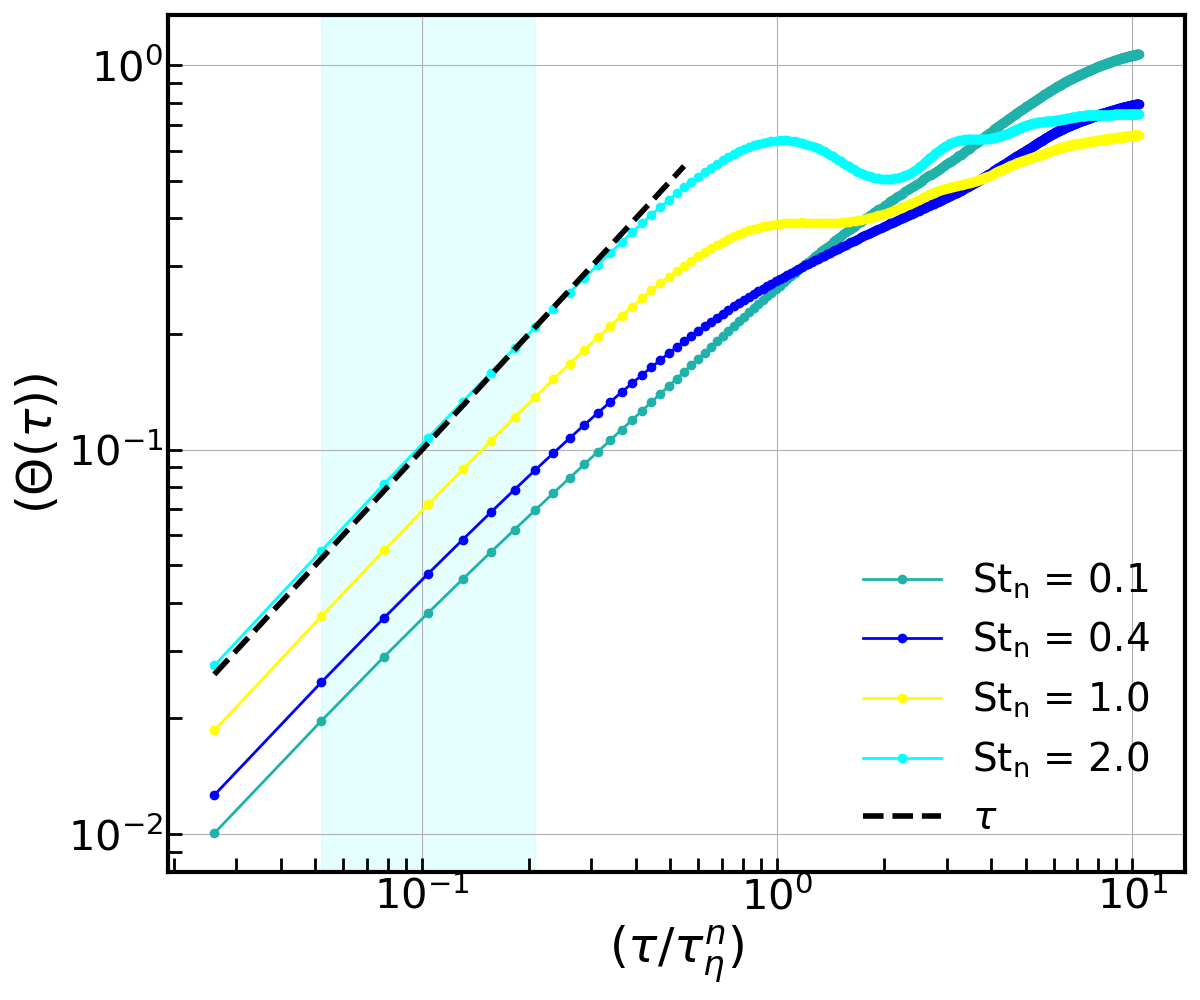}
	\put(-120,105){{\bf(b)}: T = 1.65 K}
	\put(-45,60){\tiny $\tilde{U}_{ns} = 11.32$}
	\includegraphics[width= 0.30\linewidth]{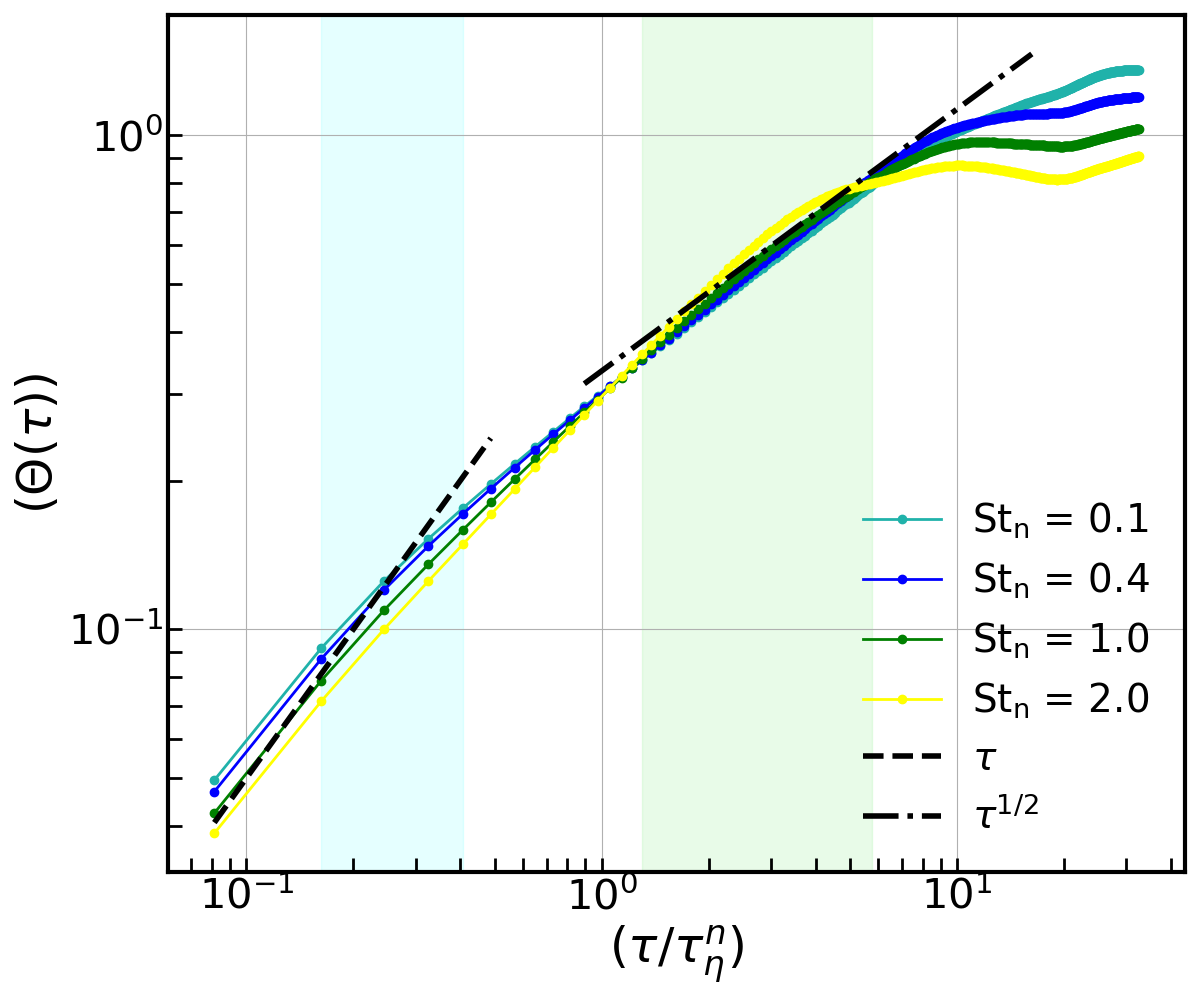} 
	\put(-120,105){{\bf(c)}: T = 2.10 K }
	\put(-40,67){\tiny $ \tilde{U}_{ns} = 8.94$}}
    \put(-180,140){\bf  Counterflow}	
	\color{black}
	\caption{Log-log plots of the angle $\Theta(\tau)$ (see text) 
		versus the time lag $\tau$ for (a) coflow ST at $T=1.65 K$, (b) and (c) counterflow ST at $T = 1.65 K$ and $T = 2.1 K$ respectively. These plots are for $\beta=1.25$ and for different values of $\rm St_n$.   
		Note the power-law scaling regions in the dissipation ranges 
		(cyan-shaded regions) and in the inertial ranges  (green-shaded 
		regions) with  $\Theta (\tau) \sim \tau^\alpha$ and 
		$\Theta (\tau) \sim \tau^\zeta$, respectively.}
\label{fig:Thetainc}
\end{figure*}

For coflow ST at $T=1.65$K [Fig.~\ref{fig:Thetainc}(a)], we present log-log plot of $\Theta(\tau)$ versus $\tau$ for different values of $\rm {St_n}$ and $\beta=1.25 (\rho_{p}/\rho o=0.7)$. These plots show two power-law scaling regions separated by a crossover regime 
around $\tau_{\eta}^n$, the Kolmogorov time scale: in the dissipation range (cyan-shaded regions) $\Theta(\tau) \sim \tau^\alpha$; in the inertial range  (green-shaded regions) $\Theta (\tau) \sim \tau^\zeta$; our data are consistent with the 
exponents $\alpha \simeq 1$ and $\zeta \simeq 1/2$. Similar scaling regimes have been obtained for Lagrangian tracers in classical-fluid turbulence~\cite{Bos2015}  except at large Stokes number in which case particles become ballistic and do not show the inertial range. This shows that coflow ST at $T=1.65$K or higher temperatures behaves like classical fluid turbulence because the normal-fluid and superfluid components are strongly coupled by the mutual friction.

 For counterflow ST at $T=1.65$K (Fig.~\ref{fig:Thetainc}(b)), the scaling region in the dissipation range ($\tau<\tau_{\eta}^n$) yields $\alpha \simeq 1$, as in coflow ST. Beyond $\tau_{\eta}^n$, because of the mean counterflow speed ($\tilde{U}_{ns}=20.31$), particles form large clusters (Fig.\ref{fig:iso_surface}(b)-(d)). For light particles (Fig.\ref{fig:iso_surface}(d)), these large clusters are attracted towards the vortex columns and are substantially confined. This confinement reduces asymptotic value of $\Theta$ at large $\tau$ (as compared to its counterpart in coflow ST). In particular, particles with large $\rm St_n$ (cyan curve in Fig.\ref{fig:Thetainc}(b)) are strongly affected by this confinement because they follow the normal-fluid component, which has large mean velocity ${\bf U}_n$ as compared to that of the superfluid component ${\bf U}_s$ [cf. Ref.~\cite{confinement} for a related effect in classical-fluid turbulence]. At a higher temperature,
say $T=2.10K$, the superfluid fraction is very small and the behavior of $\Theta$ is similar to that in classical fluid turbulence [Fig.\ref{fig:Thetainc}(c)]  with the scaling exponents $\alpha \simeq 1$ and $\zeta \simeq 1/2$; of course, at large values of $\tau$, $\Theta$ is reduced, because of the mean counterflow  velocity,
as it is for $T=1.65K$.

\subsection{Particle Trajectories}
\label{subsec:Trajectories}

In addition to the statistics of particle velocities and accelerations in coflow and counterflow ST, it is instructive to examine  the statistics of the trajectory curvature $\kappa$ and the modulus $\theta$ of the torsion. Both of these quantities have dimensions of inverse length, so large values of $\kappa$ and $\theta$ provide information about small-scale structures.
To characterize the geometry of a particle's trajectory we follow
Ref.~\cite{Akshay2014} and use the tangent $ {\bf t} $, normal ${\bf n}$, and
bi-normal $\bf b$ that are defined as~\cite{Spivak1970, Stone2009, Braun2006}
\begin{equation}
{\bf t} = \frac{d {\bf r}}{ds} ; \quad 
{\bf n} = \frac{1}{\kappa}\frac{d {\bf t}}{ds};
\quad  {\bf b} = {\bf t} \times {\bf b} .
\label{eq:tnb1}
\end{equation}
Here, $s$ is the arc length and $\kappa$ is the curvature of the trajectory;
${\bf t}, {\bf n}$, and ${\bf b}$ evolve as follows:  
\begin{equation}
\frac{d {\bf t}}{ds} = \kappa {\bf n} ; \quad
\frac{d {\bf n}}{ds} = \vartheta {\bf b} - \kappa {\bf t}; \quad
\frac{d {\bf b}}{ds} = -\vartheta {\bf n} ;
\label{eq:tnb1}
\end{equation}
$\vartheta$ is the torsion of the trajectory. In terms of ${\bf v}$ and its derivatives
($\dot {\bf v}, \ddot {\bf v} $), we have, in parametric form:
\begin{equation}
\kappa = \frac{\left|{{\bf v}\times\dot {\bf v}}\right |}
{{\left|{\bf  v}\right|}^3} = \frac{a_n}{v^2}; \quad
\vartheta = \frac{{\bf v}\cdot(\dot{\bf v}\times
	\ddot {\bf v})}{({\bf  v}\cdot {\bf v})^3 \kappa^2}. 
\label{eq:cur_tor}
\end{equation}
where $v$ and $a_n$ are the magnitude of the velocity and of the normal component of particle's acceleration.

\begin{figure*}
	\color{blue}\boxed{\color{black}
	\centering
	\includegraphics[width=0.3\linewidth]{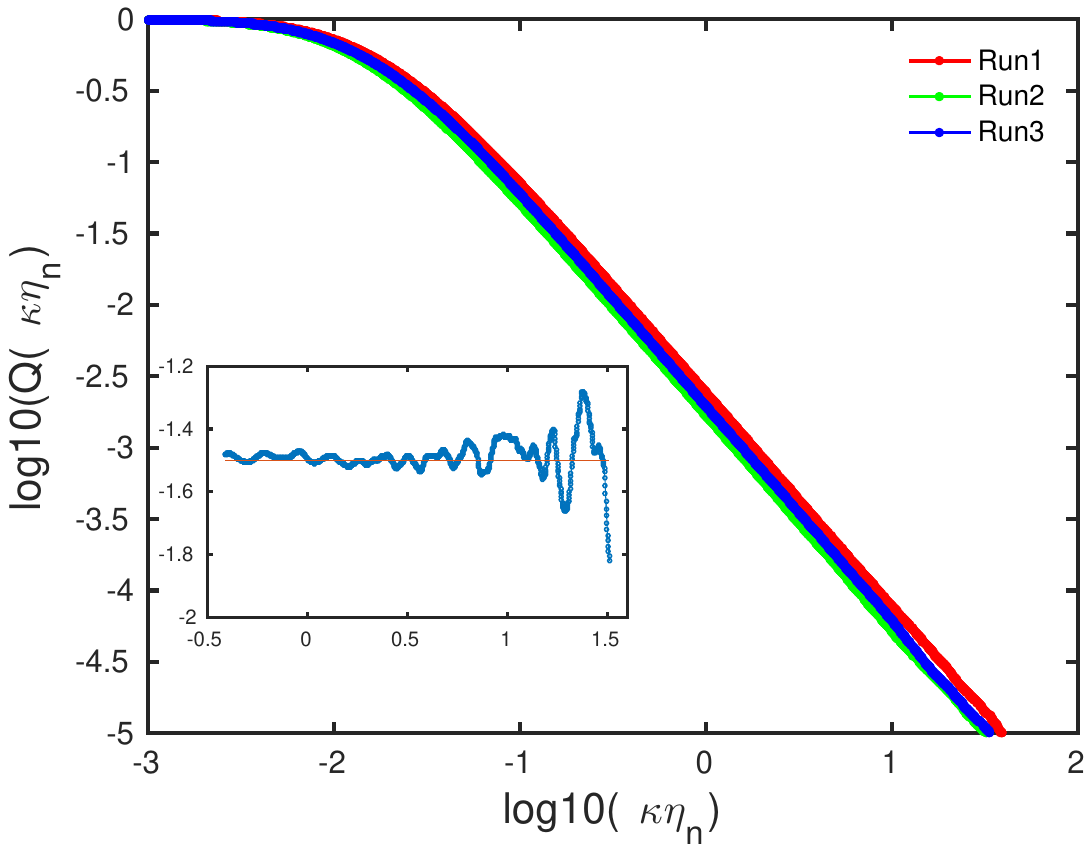} 
	\put(-130,100){\bf(a)}
	\includegraphics[width=0.3\linewidth]{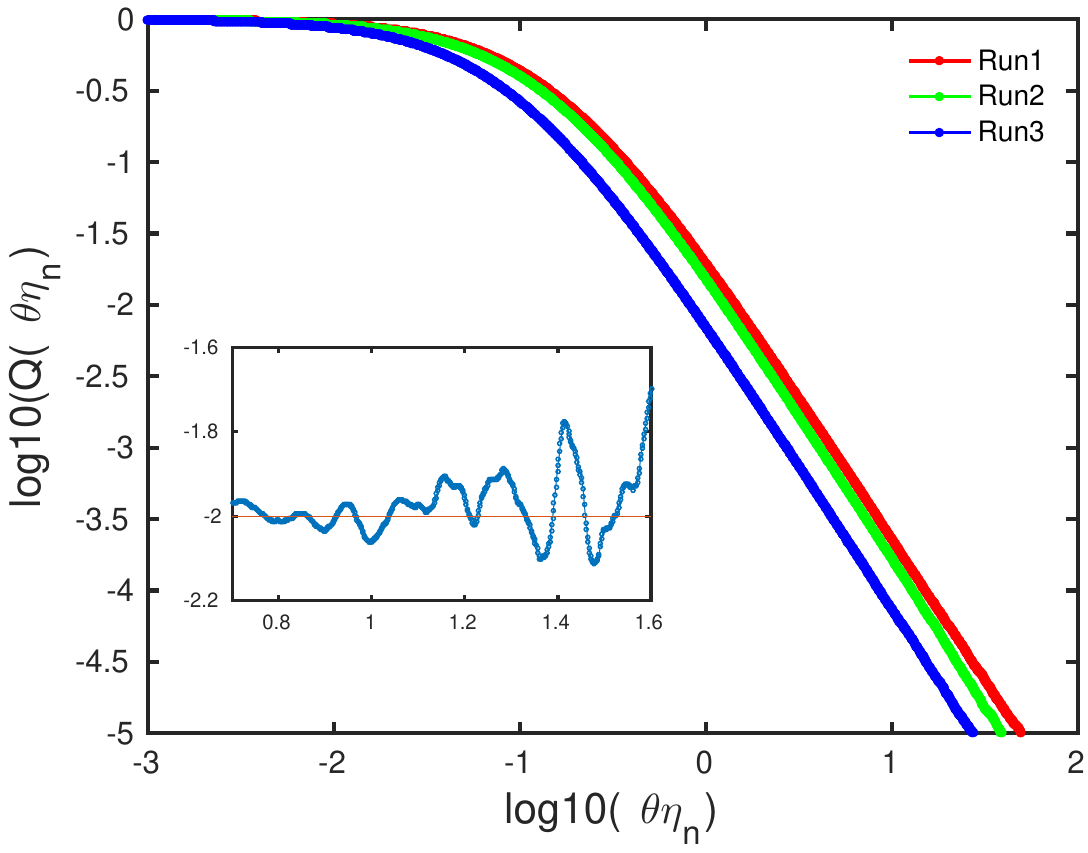} \put
	(-130,100){\bf(b)}}
    \put(-170,125){Coflow}

	\color{red}\boxed{\color{black}
	\includegraphics[width=0.3\linewidth]{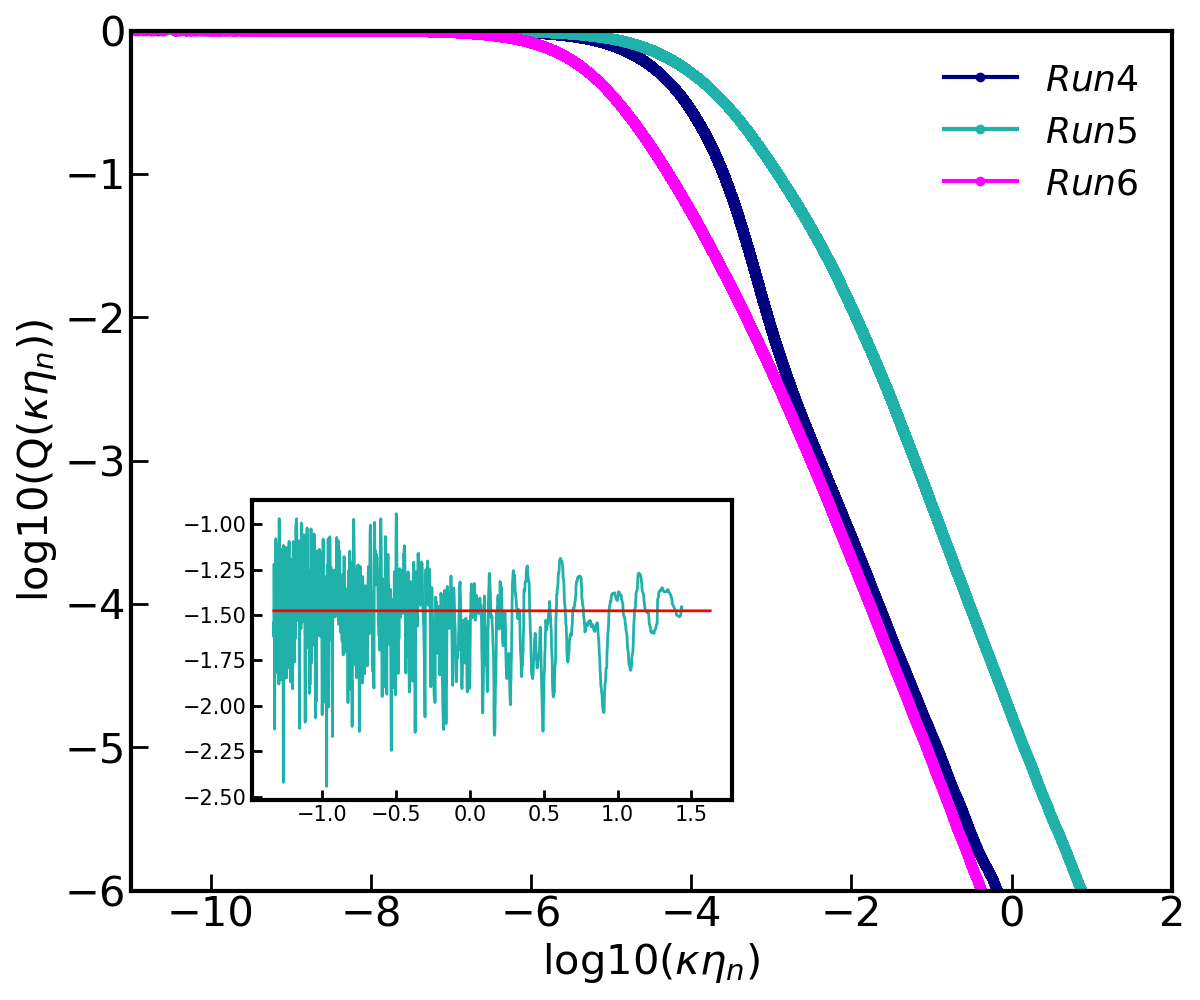}
	\put(-130,100){\bf(c)}
	\includegraphics[width=0.3\linewidth]{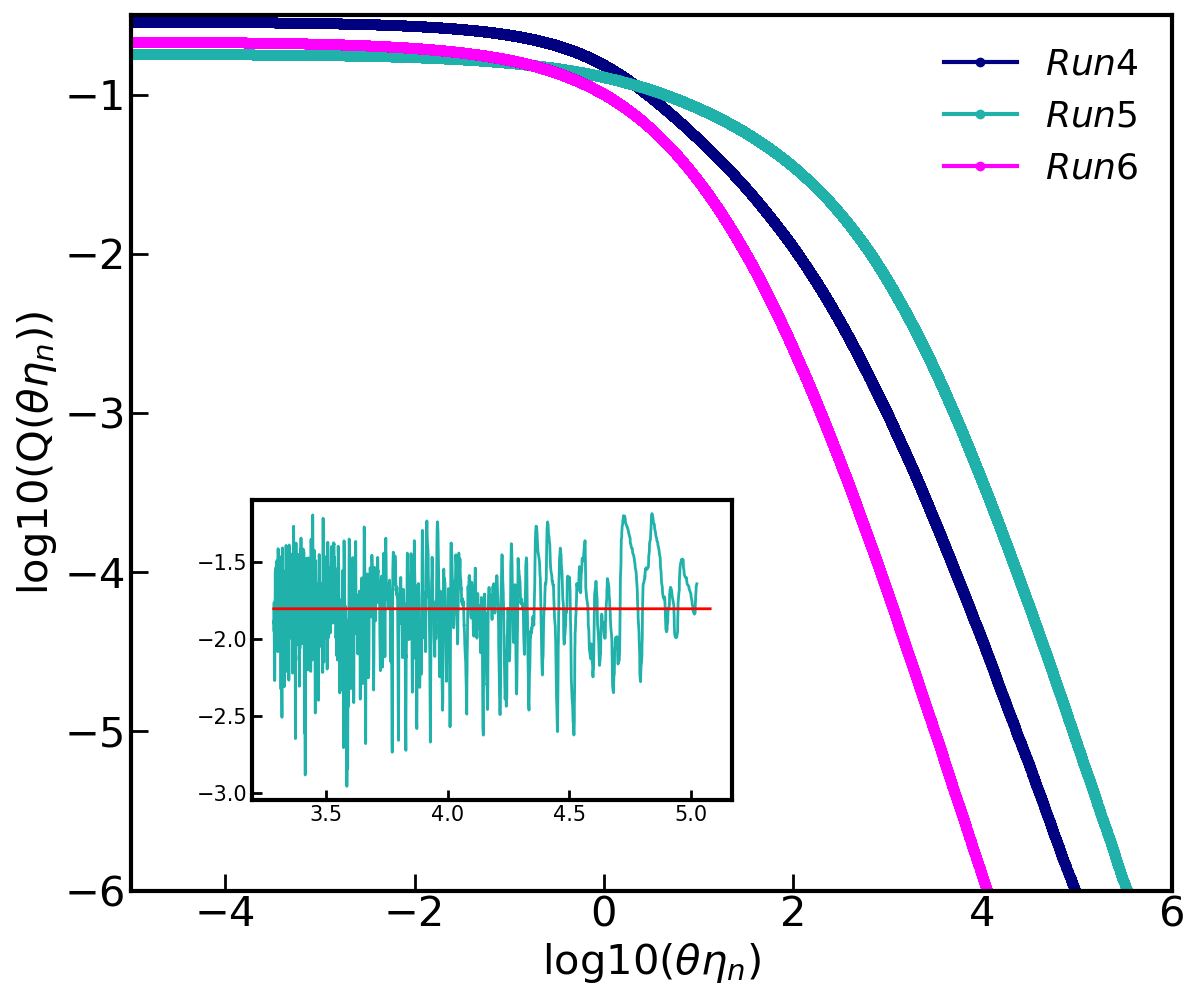}
	\put(-130,100){\bf(d)}}
    \put(-170,135){Counterflow}
	\color{black}
	\caption{Log-log plots of CPDFs of  {\bf(a)} the curvature
		$\kappa$  and {\bf(b)} the magnitude $\theta$ of the
		torsion of particle trajectories for 
		coflow ST (runs ${\bf R1}$, ${\bf R2}$, and ${\bf R3}$); 
		{\bf(c)} and {\bf(d)} are, respectively, the counterflow-ST 
		counterparts of {\bf(a)} and {\bf(b)} (for runs ${\bf R5}$, 
		${\bf R6}$).
		Insets show plots of the local slopes of the tails of these 
		CPDFs for run ${\bf R1}$ (in {\bf(a)} and {\bf(b)})
		and for run ${\bf R5}$ (in {\bf(c)} and {\bf(d)}); the mean
		values of these local slopes yield the  exponents of the 
		power-law tail of the CPDFs (and the local-slope 
		standard deviations give the error-bars for these exponents). 
		[We can also calculate the exponent of the tail of the CPDF
		of $\kappa$ from the instantaneous angles of particle 
		trajectories (see the Fig.\ref{fig:sup_append} in Supplementary
		Material \ref{Sec:supplementary}).] 
		In these plots $\rm {St_n} = 1.0$ and $\beta=0.88$.}
	\label{fig:cur_tor}
\end{figure*}

In the log-log plots of Figs.\ref{fig:cur_tor}(a) and (b), we present for
coflow ST, the CPDFs $Q(\kappa)$ and $Q(\theta)$, respectively, where $\theta =
|\vartheta|$. Both these CPDFs show power-law-scaling regions:  $Q(\kappa) \sim
\kappa^{- h_{\kappa} +1}$, for $\kappa \rightarrow \infty$, with $h_{\kappa}
\simeq 2.5$, i.e., the PDF  $ P(\kappa) \sim {\kappa}^{-h_{\kappa}} $; and
$Q(\theta) \sim \theta^{-h_{\theta} +1}$, for $\theta \rightarrow \infty$, with
$h_{\theta} \simeq 3$, i.e., the PDF $P(\theta) \sim {\theta}^{-h_{\theta}}$.
In Figs.~\ref{fig:cur_tor}(c) and (d) we present, for counterflow ST, the CPDFs
$Q(\kappa)$ and $Q(\theta)$, respectively. The exponents $h_{\kappa}$ and
$h_{\theta}$ are the same as for coflow ST.  We use a local-slope analysis
(see, e.g., Ref.~~\cite{Perlekar2011}) to calculate the mean values of
$h_{\kappa}$ and $h_{\theta}$  and their error bars (insets of
Figs.\ref{fig:cur_tor}(a) and (b)).  The exponents $h_{\kappa}$ and
$h_{\theta}$, for the tails $P(\kappa)$ and $P(\theta)$ are universal, insofar
as they are independent of $B, B', \rho_{n}, \rho_{s}, \rm {St_n}$, $\rm
{St_n}$, and $ \widetilde{ U}_{ns}$.  The exponents $h_{\kappa}$  and
$h_{\theta}$ have the same values as they do in classical-fluid
turbulence~\cite{Xu2007, Scagliarini2011, Akshay2014}.

We can obtain $h_{\kappa}$ and $h_{\theta}$, by making plausible
approximations, as in classical-fluid turbulence~\cite{Xu2007,Scagliarini2011}.
For the curvature 
\begin{equation}
P(\kappa) = \int \delta(\kappa-\frac{a_n}{v^2}) \mathcal{P}(a_n,v)da_n dv,
\label{eq:pdfkapa}	 
\end{equation}	 
where $a_n$ is the normal component of the particle's acceleration and 
$v$ the magnitude of its velocity. Furthermore,  $\theta =  \frac{|{\bf v}\cdot
	(\dot{\bf v}\times \ddot {\bf v})|} {({\bf  v}\cdot {\bf v})^3 \kappa^2}$,
which we can simplify to obtain $\theta = \frac{|\ddot {\bf v}\cdot b|}
{a_n v}$; large values of $\theta$ corresponding to small 
vaules of $a_n$ or $v$. For the modulus of the torsion 
\begin{equation}
P(\theta) = \int \delta(\theta-\frac{1}{a_n v}) \mathcal{P}(a_n,v)da_n dv.
\label{eq:pdftor}	 
\end{equation}	 
$a_n$ is a small-scale quantity and $v$ is dominated by large-scale flows,
so we argue, as in Ref.~\cite{Akshay2014}, that this scale separation suggests
\cite{Xu2007, Scagliarini2011, Akshay2014} that we have the following
factorization of the joint PDF: 
\begin{equation}
\mathcal{P}(a_n,v) \simeq P_{a_n}(a_n)P_v(v).
\label{eq:factorise}
\end{equation}
From our DNSs of the 3D HVBK model we find that:
(a) the PDF of $v$ is well approximated by the Maxwellian
\begin{equation}
P_v(v) = C_1 v^{d-1} \exp(-v^2/C_2), 
\label{eq:Maxwellian}
\end{equation}
where $C_1$ and $C_2$ do not depend on $v$; and (b) the PDF of $a_n$ can be 
fit to the form
\begin{equation}
P_{a_n}(a_n) = C_3 a_n \exp(-a_n^2/C_4),
\label{eq:PDFa_n}
\end{equation}
where $C_3$ and $C_4$ do not depend on $a_n$.  By substituting
Eqs.~(\ref{eq:factorise})-(\ref{eq:PDFa_n}) in Eqs.(\ref{eq:pdfkapa}) and
(\ref{eq:pdftor}), we get, after some simplification (in the small $a_n$
limit),
\begin{eqnarray}
P(\kappa)&\sim& \kappa^{-2.5} , \kappa \rightarrow \infty, \nonumber \\
P(\theta) &\sim& \theta^{-3},  \theta \rightarrow \infty;
\label{eq:KappaThetaScaling}
\end{eqnarray}
our DNS results are in agreement with these power-law forms.

\subsection{Energy Increments and the Irreversibility of 3D HVBK Turbulence}
\label{subsec:Irreversibility}

\begin{figure*}[!hbt]
	\color{blue}\boxed{\color{black}
		\includegraphics[width=0.25\linewidth]{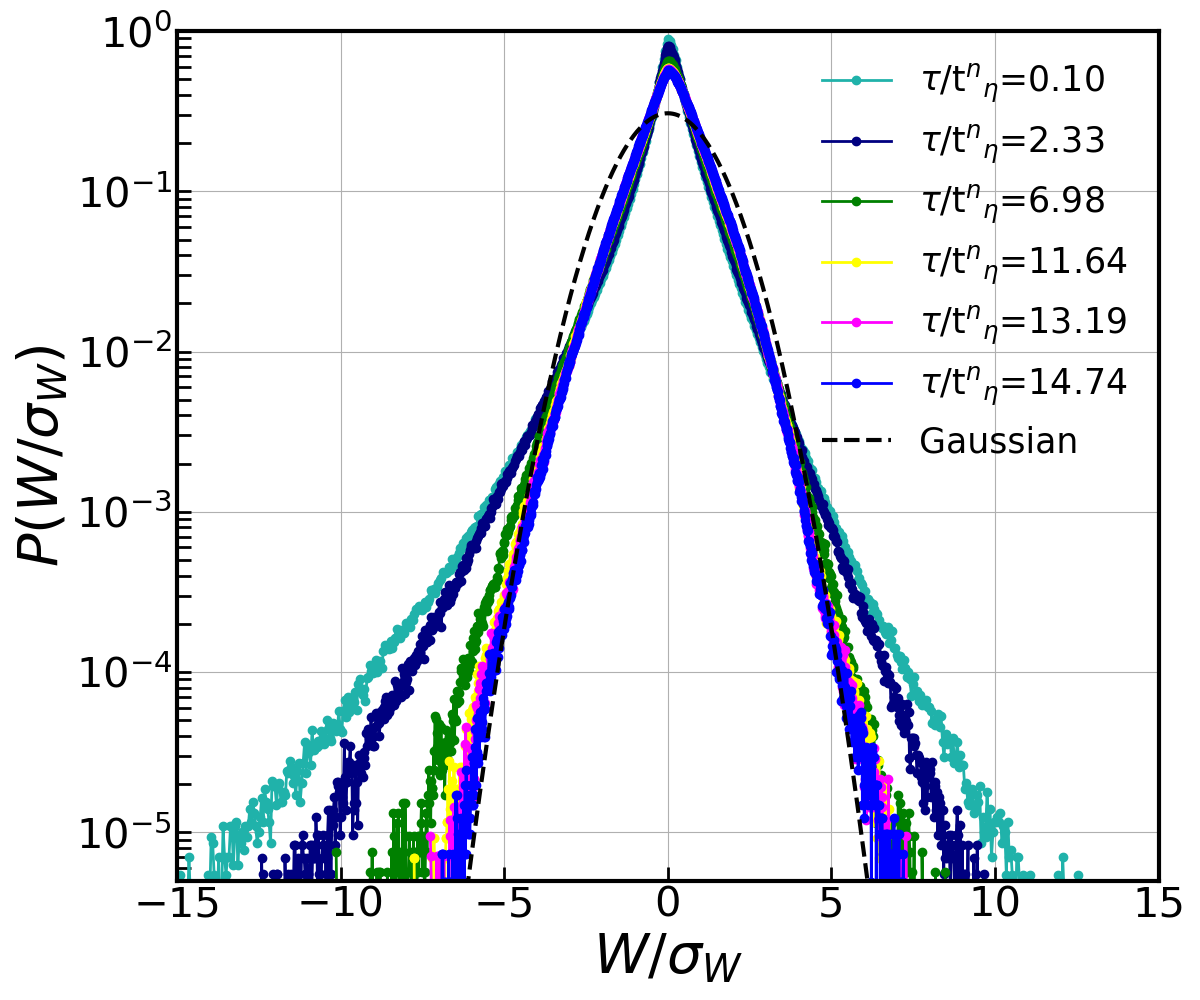} \put(-100,90){\bf(a)} 
		\includegraphics[width=0.25\linewidth]{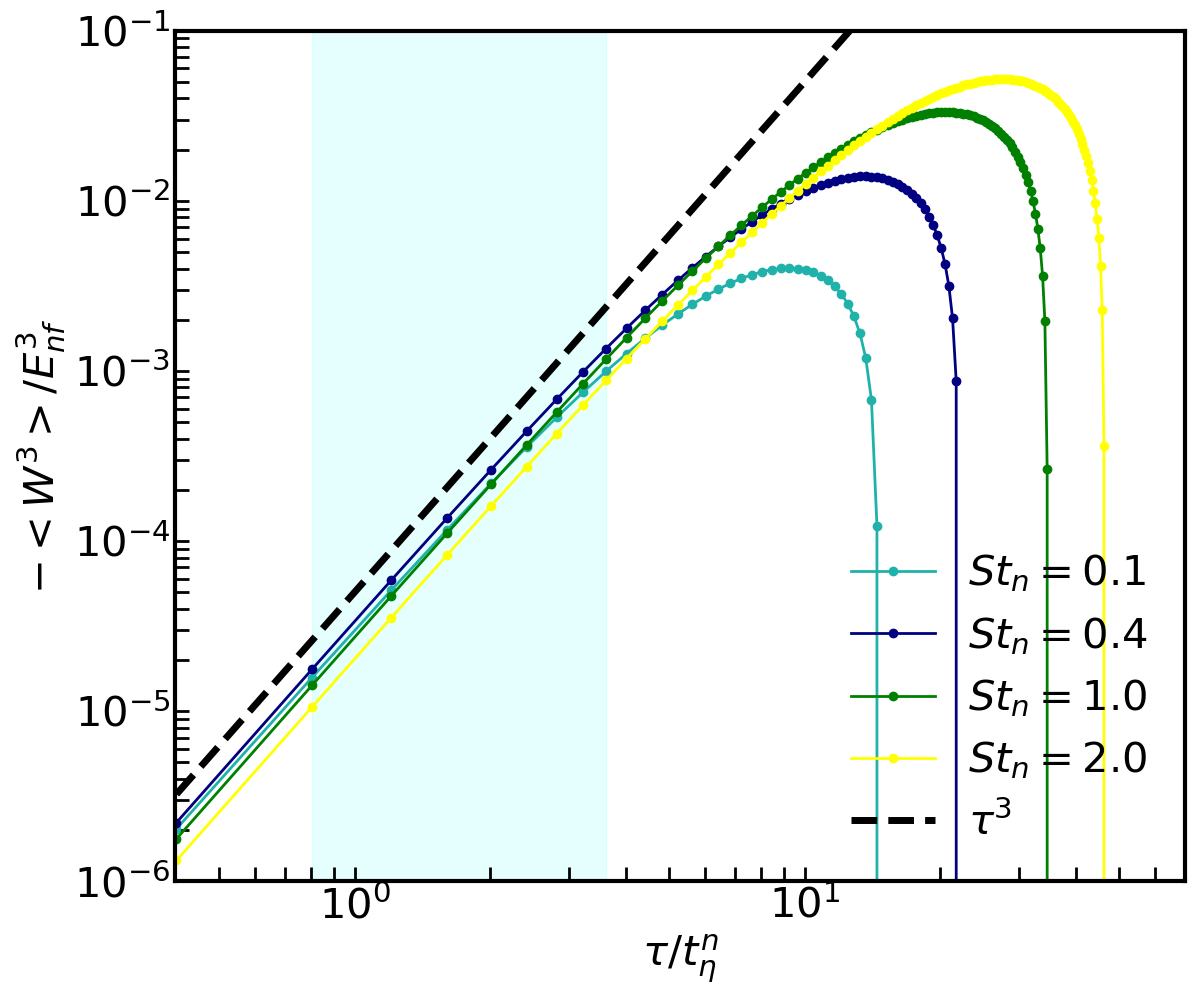} \put(-100,90){\bf(b)}
	}\put(-140,110){\bf  Coflow}
     \color{red}\boxed{\color{black}
		\includegraphics[width=0.25\linewidth]{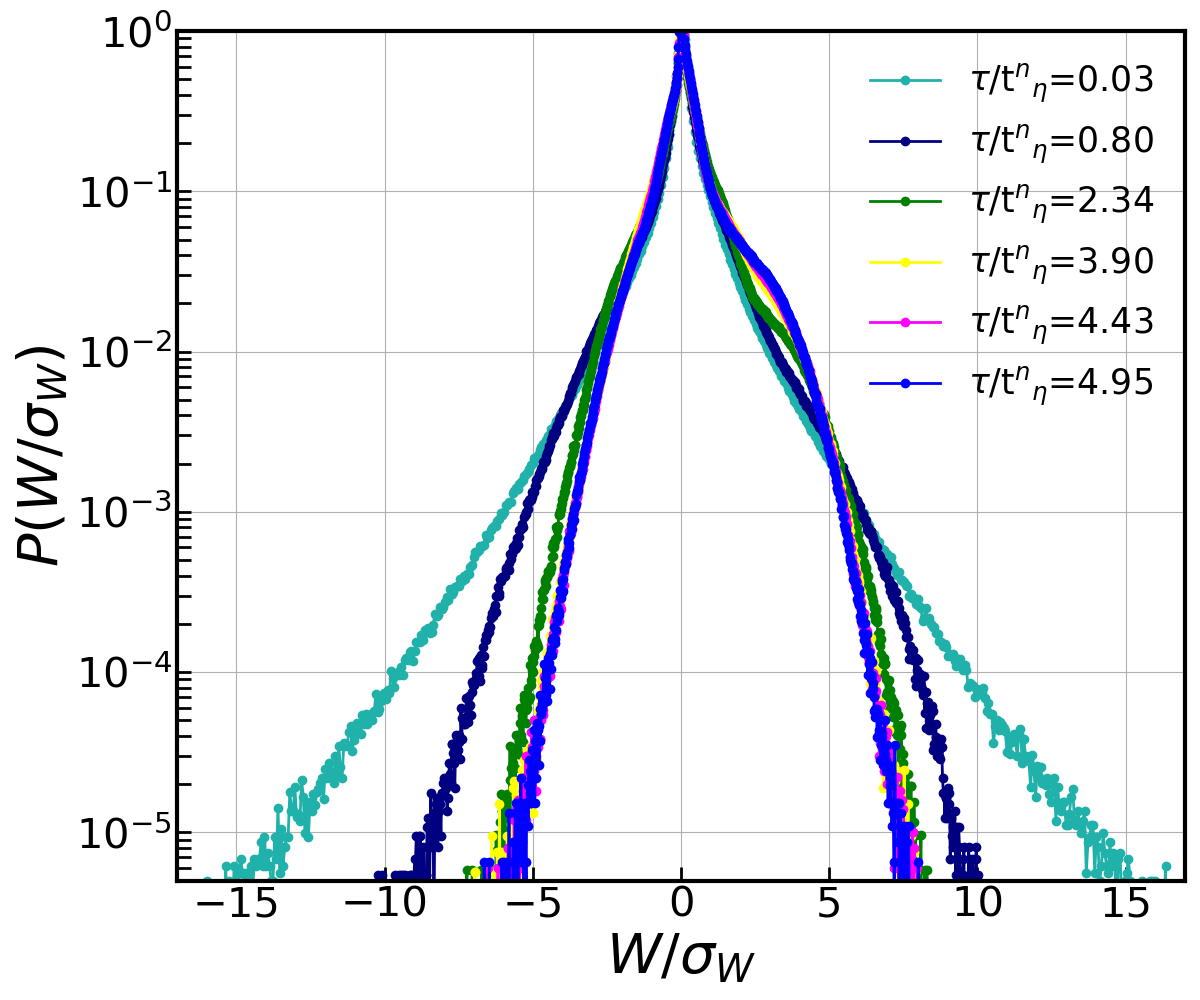}\put(-100,90){\bf(c)}	\includegraphics[width=0.25\linewidth]{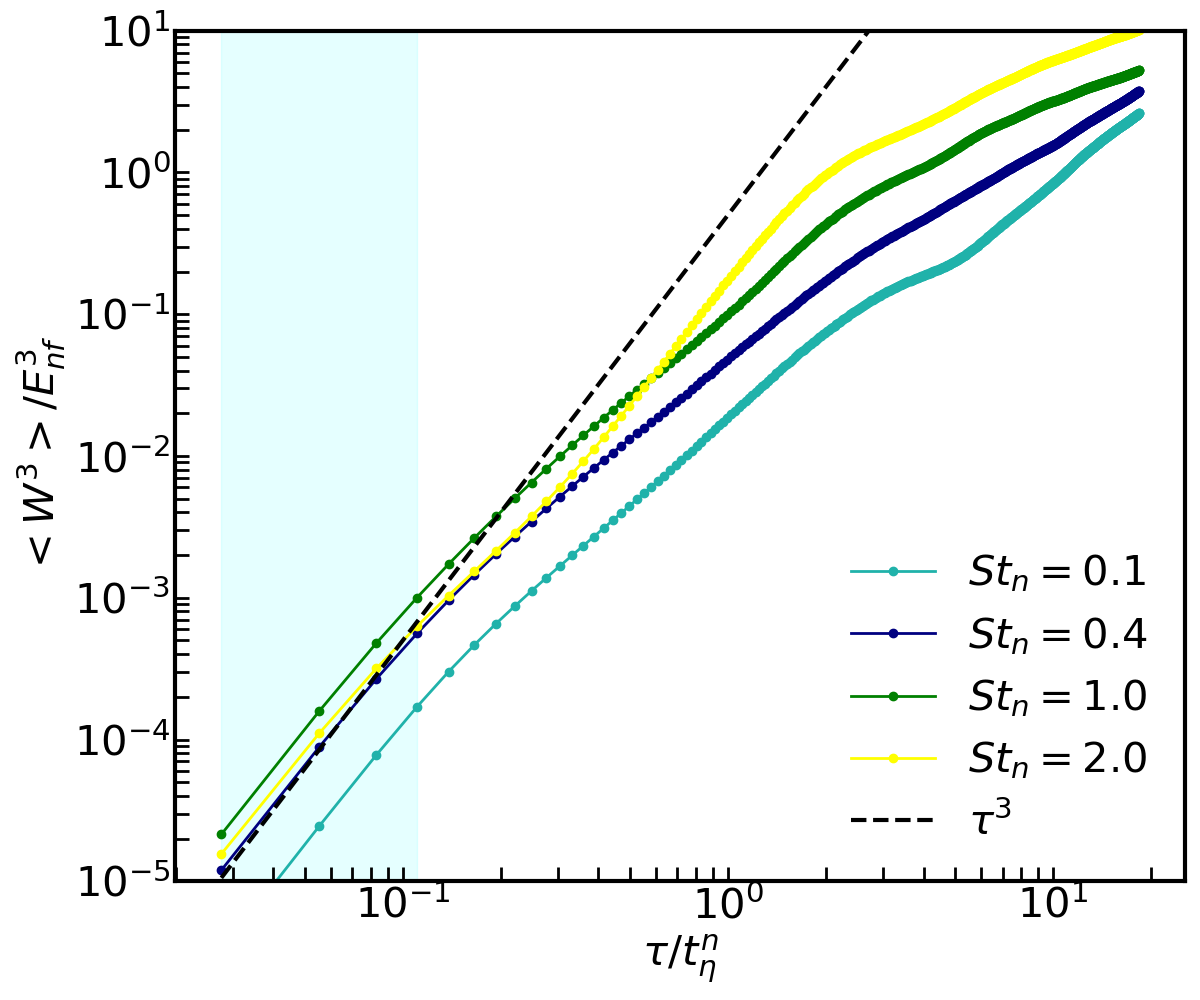}\put(-100,90){\bf(d)}
		}\put(-150,110){\bf  Counterflow}
	\color{black}
	\caption{Semilog plots of the
		energy-increment PDF ${\rm P}(W/\sigma_W)$, where $\sigma_W$ is the 
		standard deviation of $W$, for coflow ST 
		(columns 1-2) and for counterflow ST (columns 3-4) at $T=1.65K$: 
		{\bf(a)} for different time lags $\tau/t^n_{\eta}$, where $t^n_{\eta}$ is 
		the normal-fluid dissipation time; {\bf (b)} 
		log-log plots of $\langle -W^3/E^n_{\rm nf} \rangle$ versus the
		time lag $\tau/t^n_{\eta}$, where $E^n_{\rm nf}$ is the normal-fluid 
		energy, and $\tau^3$ is indicated by the black-dashed line. {\bf (c)} and {\bf (d)}, are the counterflow-ST versions of
		{\bf (a)} and {\bf (b)}, respectively. The cyan-shaded regions 
		show the regimes over which we fit power laws. The dashed curve in {\bf (a)} shows the  Gaussian fits for
		${\rm P}(W/\sigma_W)$ for $\tau/t^n_{\eta} = 14.74$ with $0$ mean and unit 
		standard deviation. For all the plots $\rm St_n=1.0$ and $\beta=0.88$.}
	\label{fig:wpdf}
\end{figure*}

We turn now to the energy increments of 
inertial particles advected by 3D HVBK turbulent flows: 
\begin{equation}
W(\tau) = \langle E_{(t_0+\tau)}- E_{(t_0)}\rangle_{t_0}, 
\label{eq:ene_inc}
\end{equation}
where $E_t = (1/2) {\bf v}(t)^2$ is the kinetic energy per unit mass of the particle and the particle velocity {\bf v} is calculated 
by using Eqs.\ref{eq:MRG}; $\langle \rangle_{t_0}$ denotes the average over the time origin $t_0$.
Such energy increments have been used to study irreversibility in
classical-fluid turbulence, where it has been found that inertial particles, in
turbulent flows of a classical fluid, gain energy slowly but lose it
rapidly~\cite{Akshay2018,Xu2014}; such gain and loss are also referred to as
flight-crash events because, on average, a particle decelerates faster than
it accelerates.  In Figs.~\ref{fig:wpdf}(a) and (c), we plot, respectively, the PDFs ${\rm P}({W}/\sigma_W)$, where $\sigma_W$ is the standard deviation, for coflow ST and counterflow ST at $T = 1.65K$ and for light particles ($\beta=1.25$).

For coflow ST, we observe that ${\rm P}({ W}/\sigma_W)$ is negatively
skewed for the small values of $\tau$, which indicates that the particles lose energy faster than they gain it. This skewness decreases as we increase $\tau$, as we show in blue curve of Fig.~\ref{fig:wpdf}(a) for coflow ST; 
clearly, these PDFs are more symmetrical (and somewhat close to Gaussian PDFs) 
than their small-$\tau$ counterparts in Fig.~\ref{fig:wpdf}(a). 

There is a striking difference if we consider light particles ($\beta=1.25$) in counterflow ST (Fig.~\ref{fig:wpdf}(c)): The skewness of ${\rm P}({W}/\sigma_W)$ is positive
(as has been found recently in a model for bacterial turbulence~\cite{bacterial_turb}). 
We conjecture that this positive skewness arises because, in counterflow ST, the mean velocity $\tilde{U}_{ns}$ makes light particles cluster near large vortical structures [Fig.\ref{fig:iso_surface}(d)].

In Figs.~\ref{fig:wpdf}(b) and \ref{fig:wpdf}(d), we present, for coflow ST and counterflow ST, respectively, and for different values of $\rm{St_n}$, graphs
of the scaled third moment of the energy increment $<{ W}^3/{E}^3_{ nf}>$ versus the scaled time increment $\tau/\tau^n_{\eta}$, where ${E}_{nf}$, and $\tau^n_{\eta}$ are, respectively, the energy and the dissipation time scale for the normal fluid. From Figs.~\ref{fig:wpdf} (b) and (d), we infer that this third moment is negative for coflow ST but positive for counterflow ST.
For small time increments in coflow ST 
\begin{equation}
-<{ W}^3/{ E}^3_{ nf}> \sim (\tau/\tau^n_{\eta})^3;
\end{equation}
and for counterflow ST
\begin{equation}
<{ W}^3/{ E}^3_{ nf}> \sim (\tau/\tau^n_{\eta})^3;
\end{equation}
deviations from these simple-scaling form are evident at large values of
$\tau/\tau^n_{\eta}$. 


Flight-crash events have also been studied  for coflow ST and
thermal-counterflow ST in experiments with superfluid $^4$He, by using
particles that are like Lagrangian tracers~\cite{Mantia2019}.  These
experiments find that, \textit{ on scales larger than the mean inter-vortex spacing and for mechanically driven coflow ST}, there are negatively
skewed PDFs ${\rm P}({ W}/\sigma_W)$, which are signatures of
flight-crash events (see above); these experimental results are in
consonance with our findings for coflow ST (Figs.~\ref{fig:wpdf}(a-b)).
Experiments~\cite{Mantia2019} have also shown that the flight-crash events are less apparent in counterflow ST than in coflow ST; and there are signatures of positively skewed velocity-difference PDFs as well; this is in agreement with our results [Figs.~\ref{fig:wpdf}(c-d)] for light particles.  Furthermore, these experiments~\cite{Mantia2019} find that,\textit{on scales smaller than or comparable to the mean inter-vortex spacing}, there is less evidence for flight-crash events than in classical-fluid turbulence; we cannot address this here because, as we have noted above, the HVBK model cannot be used for a description of superfluid turbulence on length scales smaller than or comparable to the mean inter-vortex spacing. But even in this model of HVBK, the results of counterflow are strikngly different from that of coflow.

\begin{figure}[!hbt]
	\includegraphics[width=0.5\linewidth]{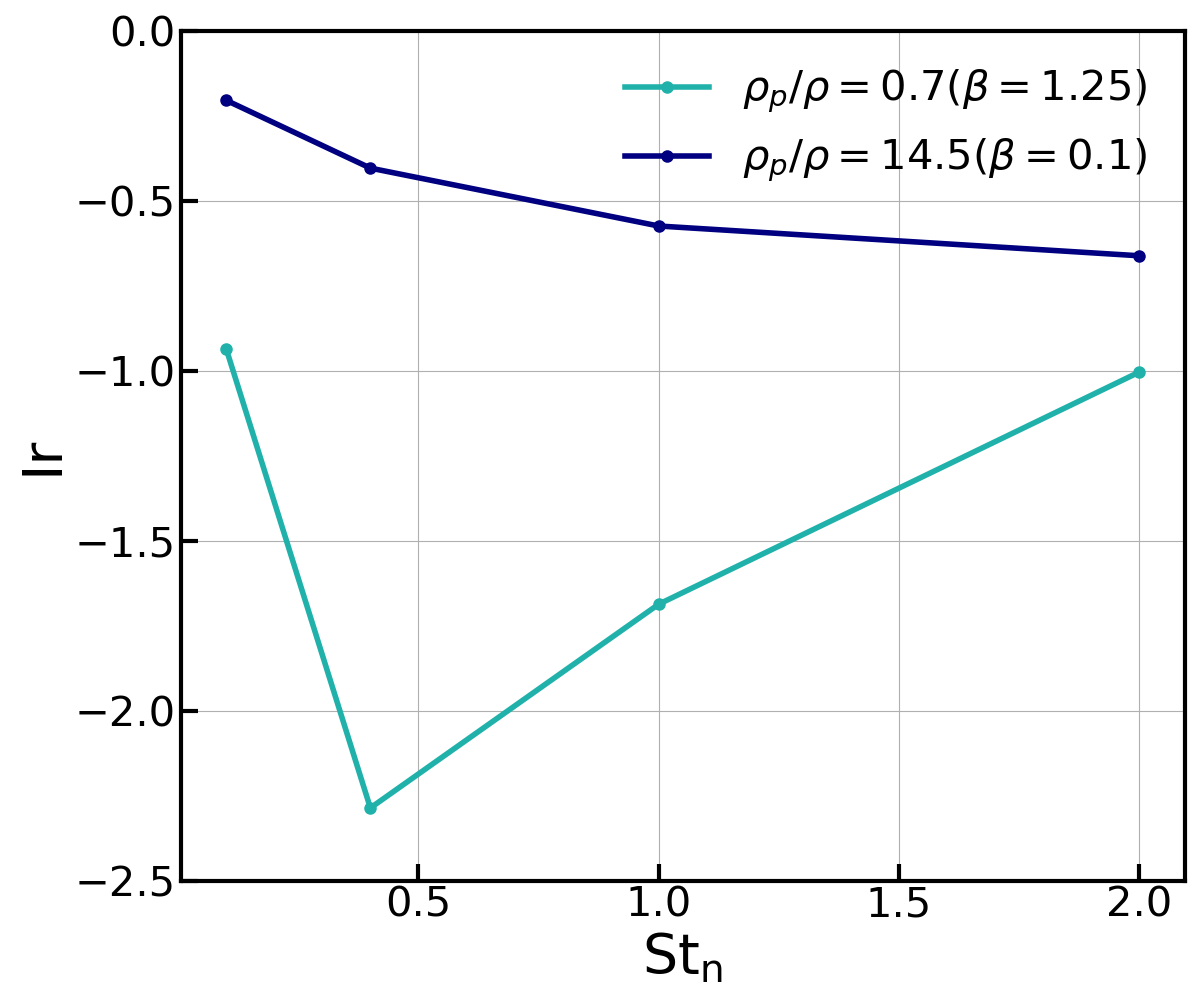}
	\put(-100, 90) {(a)}
	\put(-70, 105) {Coflow}
	\includegraphics[width=0.5\linewidth]{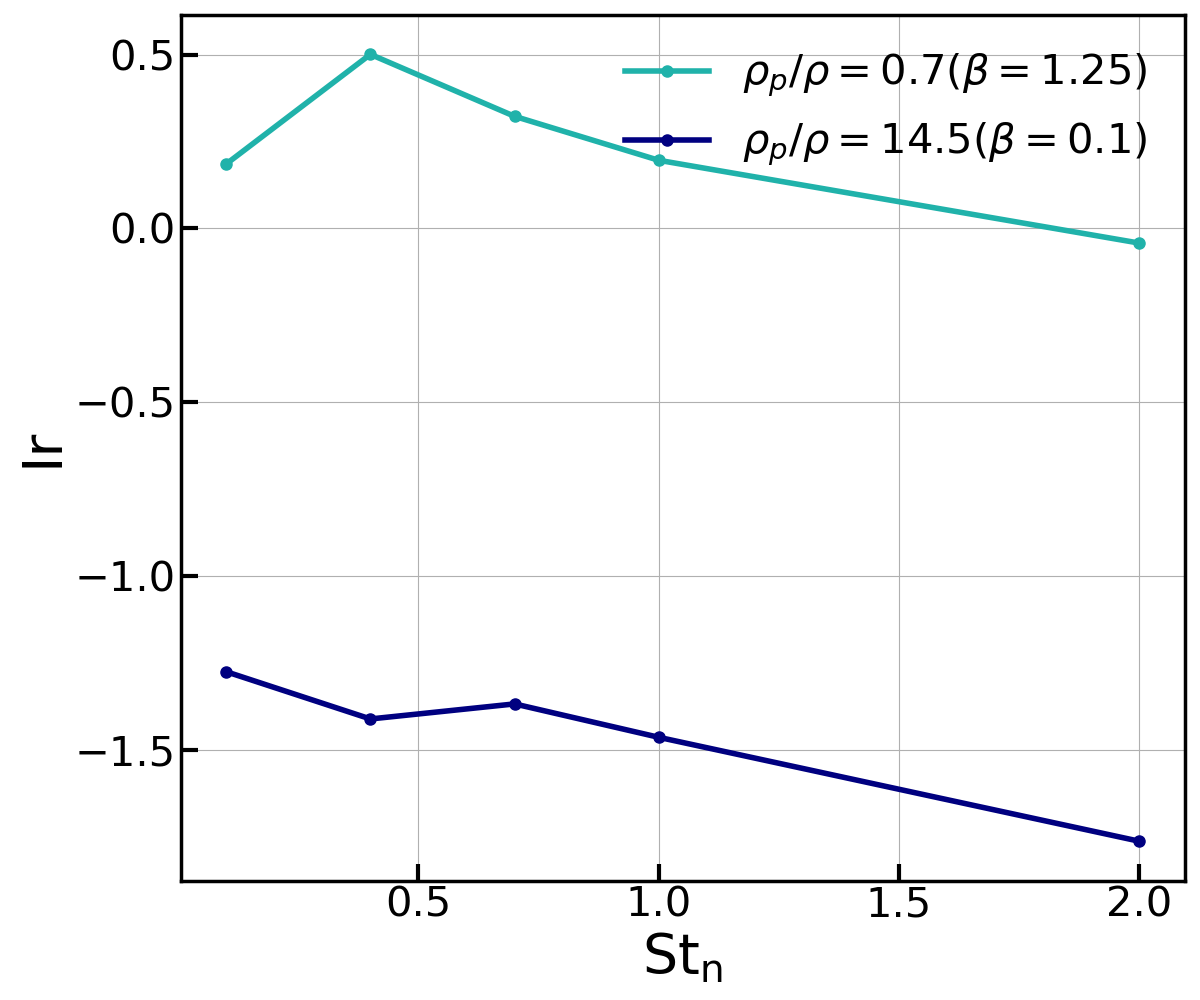}
	\put(-100, 90) {(b)}
	\put(-75, 105) {Counterflow}
	\caption{Plots of the irreversibility parameter ${\rm Ir}$ (see text)
		versus ${\rm St_n}$ for (a) coflow ST and (b) counterflow 
		ST . These plots are at $T=1.65K$ and for $\beta=0.1$ in navy color and for $\beta=1.25$ in lightseagreen.}
	\label{fig:Irr}
\end{figure}

To quantify the irreversibility of the flow, we can calculate the power $p(t) = {\bf a}(t)\cdot {\bf v}(t)$, from particle trajectories, with ${\bf a} = d{\bf v}/dt$ being the particle's acceleration. The irreversibility parameter is, as in classical-fluid turbulence~\cite{Akshay2018},
\begin{equation}
{\rm Ir} = \frac{<p^3>}{<p^2>^{3/2}},
\label{eq:irre} 
\end{equation}

which we plot versus $\rm {St_n}$ in Figs.~\ref{fig:Irr}(a) and (b) for coflow
 and counterflow ST, respectively, at $T=1.65K$ and for both light and heavy particles. For coflow ST, this irreversibility parameter is negative for light ($\beta=1.25$) as well as heavy ($\beta=0.1$) particles and for all $\rm {St_n}$; this has also been found in classical-fluid turbulence~\cite{Akshay2018}. Moreover, it has been argued~\cite{pumir2016} that ${\rm Ir} < 0$ in 3D fluid turbulence; similar arguments can be used, \textit{mutatis mutandis}, to conclude that ${\rm Ir} < 0$ in 3D HVBK coflow turbulence, in agreement with our graphs in Fig.~\ref{fig:Irr}(a). 
For counterflow ST the irreversibility parameter [Fig.\ref{fig:Irr}(b)] is positive for light particles ($\beta=1.25$), which reflects the positive skewness in the energy increments of Fig.\ref{fig:wpdf}(c)-(d); in contrast, for heavy particles ($\beta=0.1$), the irreversibility parameter is negative [navy-blue curve in Fig.\ref{fig:Irr}(b)], which indicates negatively skewed PDFs of energy increments.

\section{Conclusions} 
\label{Sec:Conclusions}
Studies of inertial particles in superfluid turbulence are in their
infancy; by contrast, there have been extensive studies of
the statistical properties of such particles advected by classical-fluid
turbulence~\cite{Toschi2009, Bec2005}. Hence, we have carried out a systematic
study of inertial particles in statistically steady coflow ST and counterflow
ST in the 3D HVBK model, for different values of the Stokes numbers $\rm
{St_n}$, with normal-fluid fractions and mutual-friction
coefficients that are taken from measurements~\cite{Donnelly1998} on superfluid
$^4$He, as a function of the temperature.  One recent study~\cite{Giorgio2020}
has investigated the clustering of inertial particles in 3D HVBK turbuence and
has shown that, for coflow ST, although the particle distribution is nearly
uniform at high temperatures, it still has signatures of some clustering. 

Coflow ST is isotropic but counterflow ST is inherently anisotroic; we have shown this via isosurfaces of $|\omega_n|$ and the positions of representative particles in Fig.~\ref{fig:iso_surface}. For coflow ST at $T=1.65K$, particles cluster  as they do in classical-fluid turbulence because, at this temperature, the mutual friction couples both fluids strongly. The particles form large-scale clusters at $T=1.65K$ in counterflow ST; and light particles are attracted towards [Fig.~\ref{fig:iso_surface} (d)] the large vortical columns; by contrast, heavy particles are expelled from these vortical columns  [Fig.~\ref{fig:iso_surface}(c)].

These large vortical columns have a direct influence on the statistics of the angle $\Theta$, which is the angle between subsequent inertial-particle-displacement increment. The study of $\Theta$ reveals two scaling regions; one in dissipation and other in the inertial region. In case of coflow ST, the large time asymptotic value of $\Theta$ is the same for all Stokes number which is the signature of isotropic case~\cite{Bos2015}. While for counterflow this asymptotic value of $\Theta$ reduces for light particles with large $\rm St_n$ as they are affected more by the confinement from normal fluid component. Ref~\cite{confinement} studies the effect of mean velocity on the angle $\Theta$ in case of classical turbulence and also observe such reduction in the large time lag value of $\Theta$.

One of the main results of this study is the signature of positive skewness in the PDFs of energy increment~\ref{fig:wpdf}(c)-(d) for light particles. As we mention earlier, in a recent study of coflow and counterflow ST, Ref~\cite{Mantia2019} observe that the flight crash events are less prominent than that of classical fluid turbulence; they show that for coflow there is some similarity to classical case at large length scales. This result of coflow is in agreement with our study .i.e there are signatures of flight crash events in HVBK model of coflow. For counterflow, Ref~\cite{Mantia2019} observe different results from the classical case at all length scales and found signatures of positive skewness in moments of velocity differences. This is also in consonance with our results of positive skewness in case of counterflow for light particles; while for heavy particles the PDFs of energy increment~\ref{fig:wpdf} are negatively skewed. 

We hope that our definition and study of flight crash events, for
inertial particles in 3D HVBK turbulence, will lead to new experimental
investigations of this problem in, e.g., superfluid $^4$He or Bose-Einstein
condensates (BECs).

\section{Supplementary Material}
\label{Sec:supplementary}

In the Supplementary Material, we provide: (1) a brief description of the specific power laws found in Fig.~\ref{fig:aniso_spectra}; (2) iso-surface plots of the magnitude of the normal-fluid vorticity $|\omega_n|$ at temperature $T=2.10K$; (3) CPDFs of the persistence time~\ref{subsec:Persistence}, $t_s^{per}$, at $T=1.65K$ for the superfluid component; (4) the curvature~\ref{subsec:Trajectories}, $\kappa$, of particle trajectories, obtained from the instantaneous angle $\Theta(t,\tau)$;  (5) iso-surface plots of the magnitude of the normal-fluid vorticity $|\omega_n|$ at temperature $T=2.10K$ for a square cuboid domain with resolution $256\times 256\times 1024$.
\onecolumngrid
\section{Appendix}

\begin{figure}[!hbt]
	\includegraphics[width=0.5\linewidth]{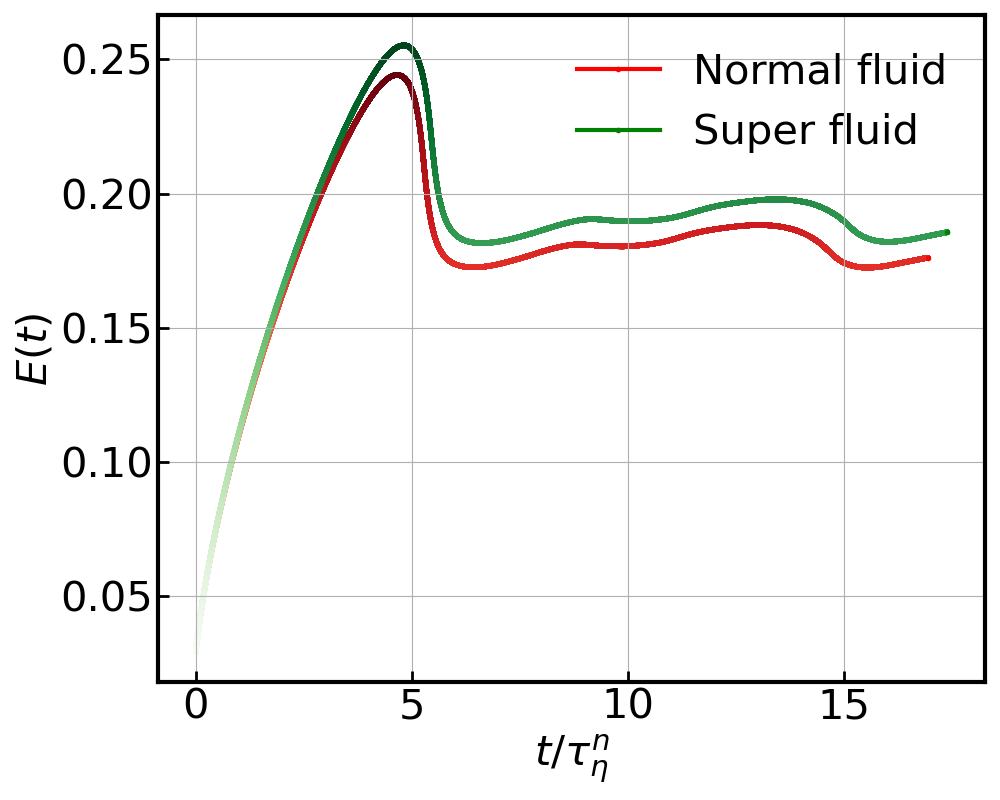}
	\caption{Plot of the volume-averaged energy, $E(t)= \sum_{k_{\parallel}}E_{k_{\parallel}} + \sum_{k_{\perp}}E_{k_{\perp}}$, of the turbulent flow for normal fluid and super-fluid components showing the statistically steady state at long times. Here $E_{k_{\parallel}}$ and $E_{k_{\perp}}$ are defined in Eqs.\ref{eq:spectra} and $\tau_{\eta}^n$ is the dissipation time scale for the normal fluid. }
	\label{fig:steady_kinetic}
\end{figure}

\begin{table}[!hbt]
	\centering
	\begin{tabular}{|c|c|}
		\hline
		{\bf Variable} & {\bf Description}\\
		\hline
		$\tau_{\eta}^{n/s}$ & Kolmogorov-dissipation time scale for normal fluid/superfluid\\
		\hline
		${\bf U_{ns}} (\widetilde{U}_{ns})$ & Counterflow mean velocity (magnitude of mean relative velocity) \\
		\hline
		$\beta$ & Parameter that accounts for the added mass effect to the particle\\
		\hline
		$\rho_p$ & Particle's density\\
		\hline
		$\Theta$ & Angle between particle's subsequent position increment\\
           \hline
           $\alpha, \zeta$ & exponents of the angle $\Theta$ \\
		\hline
		$\kappa$ &  Curvature of the particle's trajectory\\
		\hline
		$\theta$ & Magnitude of the torsion of the particle's trajectory\\
		\hline
		$Q_{n/s},R_{n/s}$ & Invariants of velocity gradient tensor for normal-fluid/superfluid\\
		\hline
		$t^{per}_{n/s}$ & Persistence time of particles for normal-fluid/superfluid\\
		\hline
		$W(\tau)$ & Particle's kinetic energy increment separated by time lag $\tau$ \\
		\hline
		$p$ & Power input to the particle \\
		\hline
		Ir  & Irreversibility \\
		\hline

	\end{tabular}
\caption{\label{tableX} Table consisting of all the variables used in the 
paper along with their descriptions.}
\end{table}
\twocolumngrid

\section*{Acknowledgments} 
\label{Sec:Conclusions}

We thank Samriddhi Sankar Ray and Kiran Kolluru for discussions, SERB and CSIR
(India) for support, and the National Supercomputing Mission (NSM) and SERC
(IISc) for computational resources.  SS acknowledges support from the PMRF. VS
acknowledges support from the Start-up Research Grant No.  SRG/2020/000993 from
SERB, India, Grant No.  IIT/SRIC/ISIRD/2021-2022/03 from the Institute Scheme
for Innovative Research and Development (ISIRD), IIT Kharagpur, and the NSM for
providing computing resources of ‘PARAM Shakti’ at IIT Kharagpur, which is
implemented by C-DAC and supported by the Ministry of Electronics and
Information Technology (MeitY) and Department of Science and Technology (DST),
Government of India. AKV and SS contributed equally to this study.

\section*{AUTHOR DECLARATIONS}
\subsection*{Conflict of Interest}
The authors have no conflicts to disclose.

\section*{Data Availability}

The data that support the findings of this study are available from corresponding author upon reasonable request.

\onecolumngrid

\section{Supplementary Material}
\renewcommand{\thefigure}{S\arabic{figure}}

In this Supplemental Material, we provide the following:

\begin{enumerate}
	
	\item {{\bf Specific powers laws in counterflow}: These specific powers are
		standard in the study of statistically homogeneous isotropic turbulence
		in the phenomenology of Kolmogorov-1941 (K41) theory. At the
		level of K41, the energy spectrum $E(k)\sim k^{-5/3}$; this
		energy spectrum is obtained when we average over spherical
		shells in $k$-space (by using an isotropic version of Eq. (9)
		in the main paper).  Given the inherent anisotropy of
		counterflow ST, the energy spectrum is a function of
		$k_{\parallel}$ and $ k_{\perp}$, which are the wavevectors
		along the counterflow direction and perpendicular to it,
		respectively (see the definitions below Eq. (9) in the main
		paper). In the perpendicular plane, we average over circular
		shells (the second line of Eq. (9)); this yields an energy
		spectrum $\sim k_\perp^{-8/3}$ (at the level of K41 arguments);
		similarly, in the one-dimensional parallel direction, the
		energy spectrum $\sim k_{\parallel}^{-11/3}$ (at the level of
		K41 arguments); the spectral exponent is $-5/3$, for a
		spherical average, $-8/3$, for a circular average, and $-11/3$
		along one direction (at the level of K41 arguments). This is in
		agreement with the recent results of Ref. [36] as we have
		mentioned clearly in the main paper.}

	\item {In Fig.~\ref{fig:sup_iso} we show the isosurface plots of the magnitude of
		the super-fluid vorticity $|\omega_s|$ at $T=2.10K$ for counterflow ST
		and $\widehat{\bf U}_{ns} =\widehat{e}_k$ (run {\bf R6}). }
	
	\begin{figure*}[!hbt]
		\includegraphics[width=0.28\linewidth]{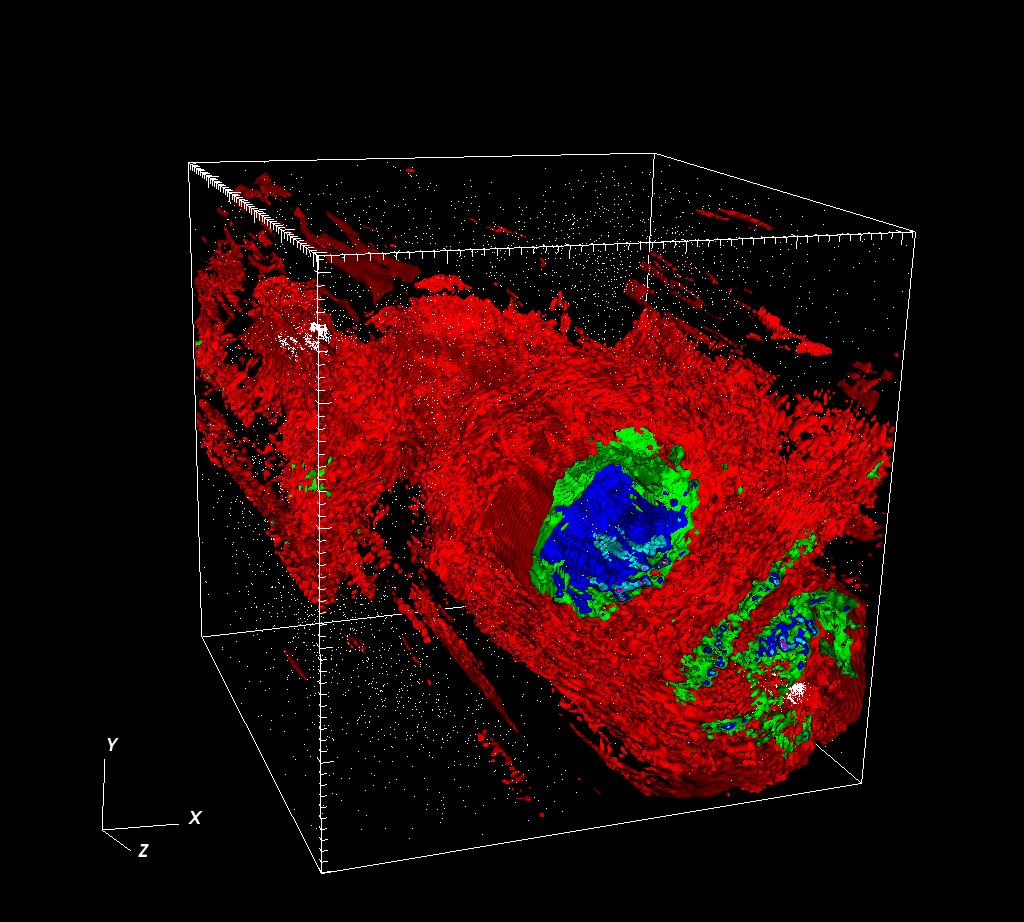}
		\includegraphics[width=0.28\linewidth]{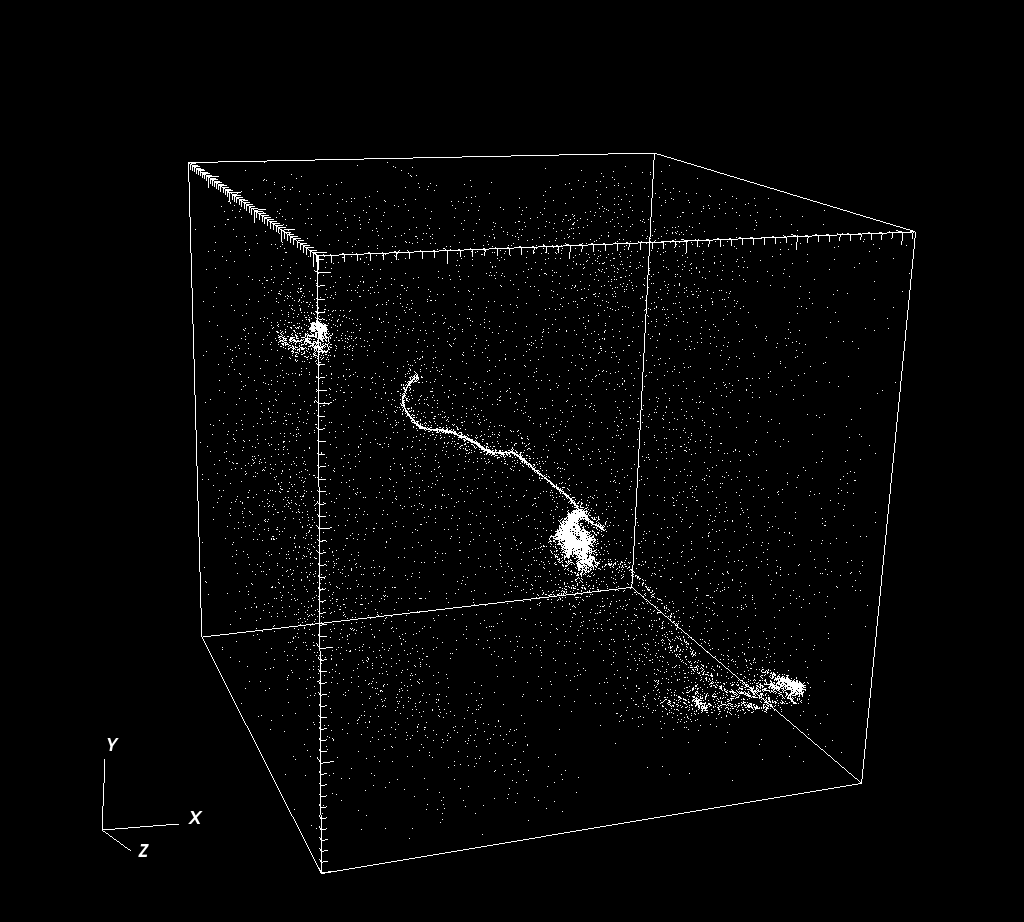}
		\put(-100,120){$\rm \color{white}\beta = 1.25$}
		\includegraphics[width=0.28\linewidth]{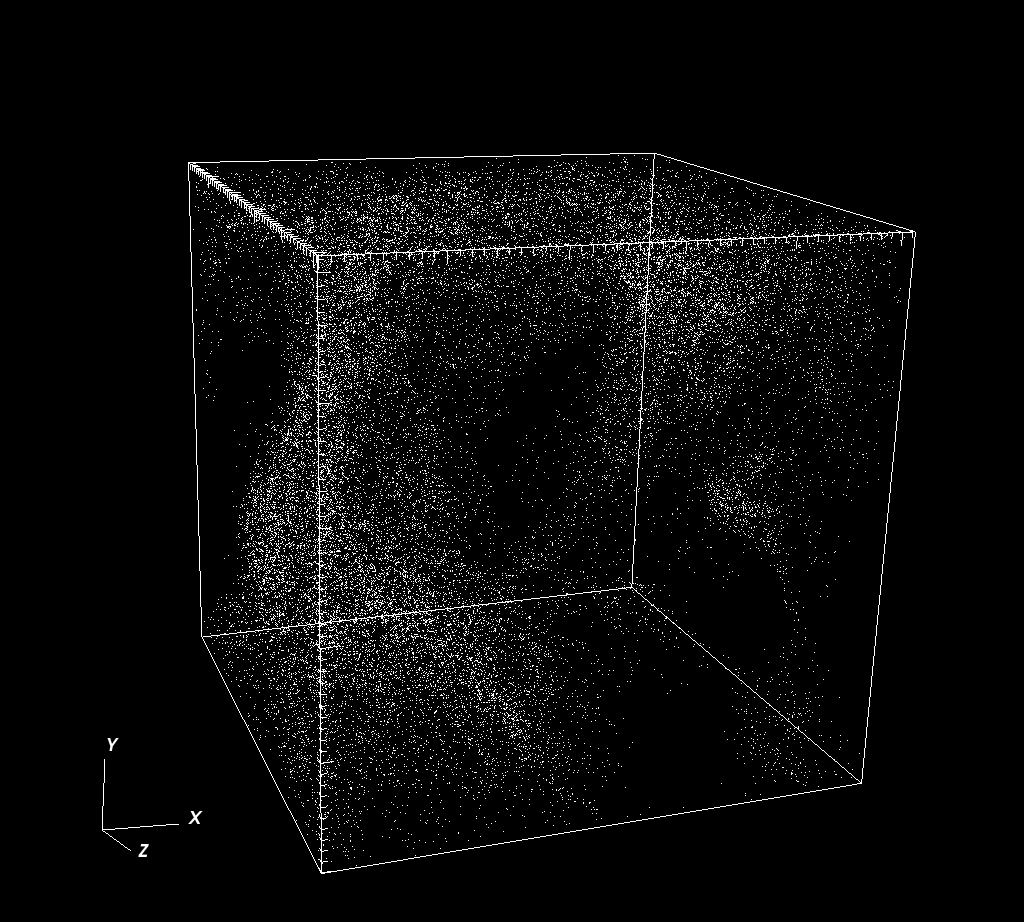}
		\put(-100,120){$\rm \color{white}\beta = 0.1$}
		\caption{Isosurface plots of the magnitude of the super-fluid vorticity 
			$|\omega_s|$, for counterflow ST ,
			$\widehat{\bf U}_{ns} = \hat{e_k}$, and $T=2.10$K. We indicate by small white spheres the positions of particles (with $\rm St_n = 1.0 $). Column2 and column3 show the positions of light ($\beta=1.25$) and heavy ($\beta=0.1$) particles respectively.}
		\label{fig:sup_iso}
	\end{figure*}

\item We plot the cumulative probability distribution functions (CPDFs) of the persistence times (see
the main paper) in the Eulerian frame for the superfluid component; for coflow ST in Fig.\ref{fig:sup_eul_per}(a) and for counterflow ST in Fig.\ref{fig:sup_eul_per}(b).  \\

\begin{figure}
	\includegraphics[width=0.4\linewidth]{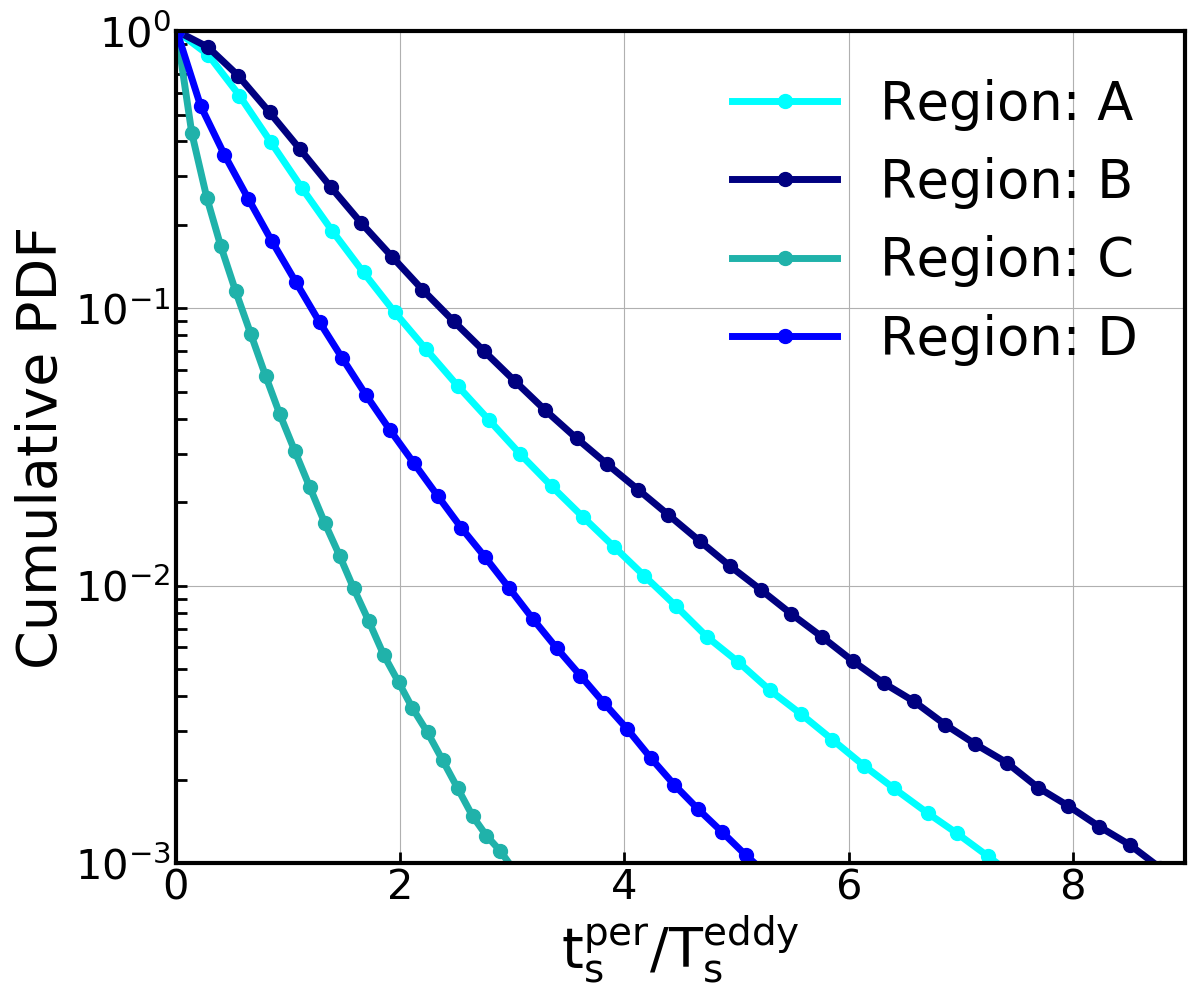}
	\put(-140,150){ (a)}
	\put(-110,170){ {\bf Coflow}}
	\includegraphics[width=0.4\linewidth]{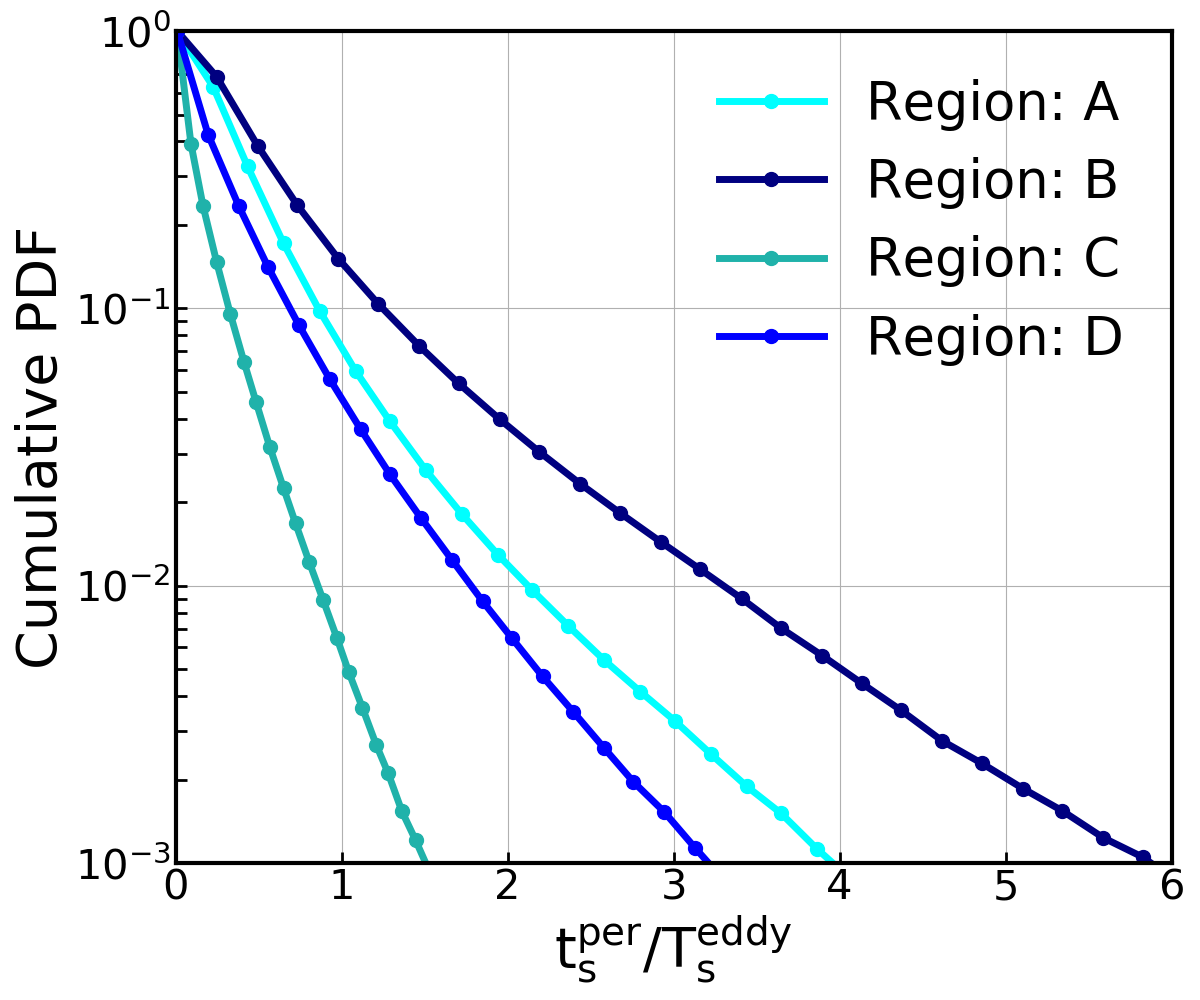}
	\put(-140,150){ (b)}
	\put(-120,170){ {\bf Counterflow}}
	\caption{Semilog plots of the CPDFs of the persistence times, $t^{per}_n$, in the  Eulerian frame for the superfluid ($s$) component; for coflow ST in (a) and for counterflow ST in (b).}
	\label{fig:sup_eul_per}
\end{figure}

\newpage

\end{enumerate}
\section{Appendix}

We can also calculate the curvature of particle trajectories from the 
instantaneous angle $\Theta(t,\tau)$. The curvature of particle
trajectories at time t for small time lag $\tau$ in terms of 
instantaneous angle $\Theta(t,\tau)$ \cite{Bos2015}is 
\begin{equation}
\kappa(t) = \lim_{\tau \to 0} \frac{\left|\Theta(t,\tau)\right|}
{2\tau \|v(t)\|}. 
\end{equation}
Log-log plots of these CPDFs are shown in Fig.~\ref{fig:append}. The slope of the tail of CPDFs are 2.5 and it is same as
calculated in the main part of this article. Here, We have calculated the slope
of curvature in y-range from -2.0 to -4.5.\\

\begin{figure}[!hbt]
\includegraphics[width=0.5\linewidth]{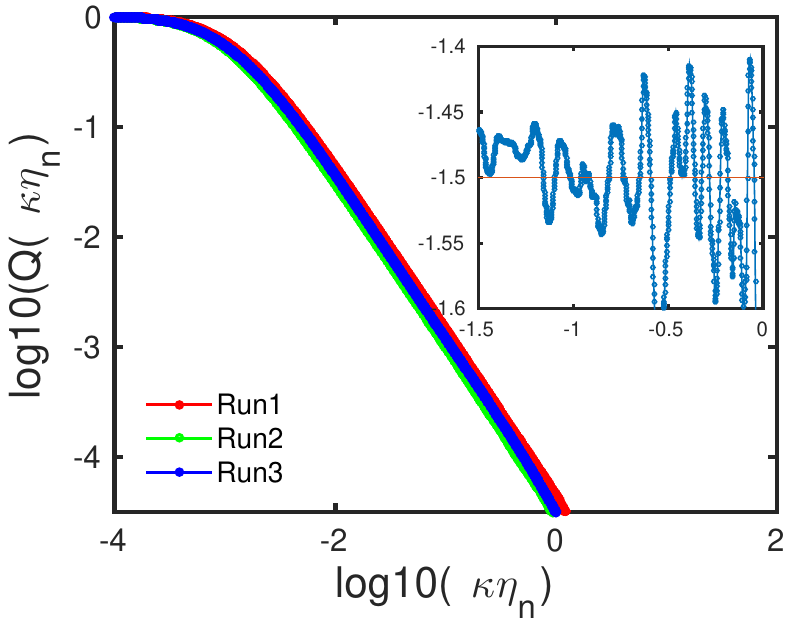} 
\caption{Log-log plots of CPDFs of the curvature
	$\kappa$ of particle  trajectories, calculated from the  instantaneous
	angle $\Theta(t,\tau)$, at time $t$ for the time lag $\tau$ from runs 
	$\bf {R1}$, $\bf {R2}$, and ${\bf R3}$. The inset shows a plot of the
	the local slopes of the tail of the CPDF for $\bf {R1}$; the mean
	value of these local slopes gives the  exponent of the power-law tail
	of the CPDF and the standard deviations are the error-bars for
	the exponents. These above plots are for $\rm {St_n} = 0.1$.}
\label{fig:sup_append}
\end{figure}

\begin{figure}
\centering
\includegraphics[width=0.5\linewidth]{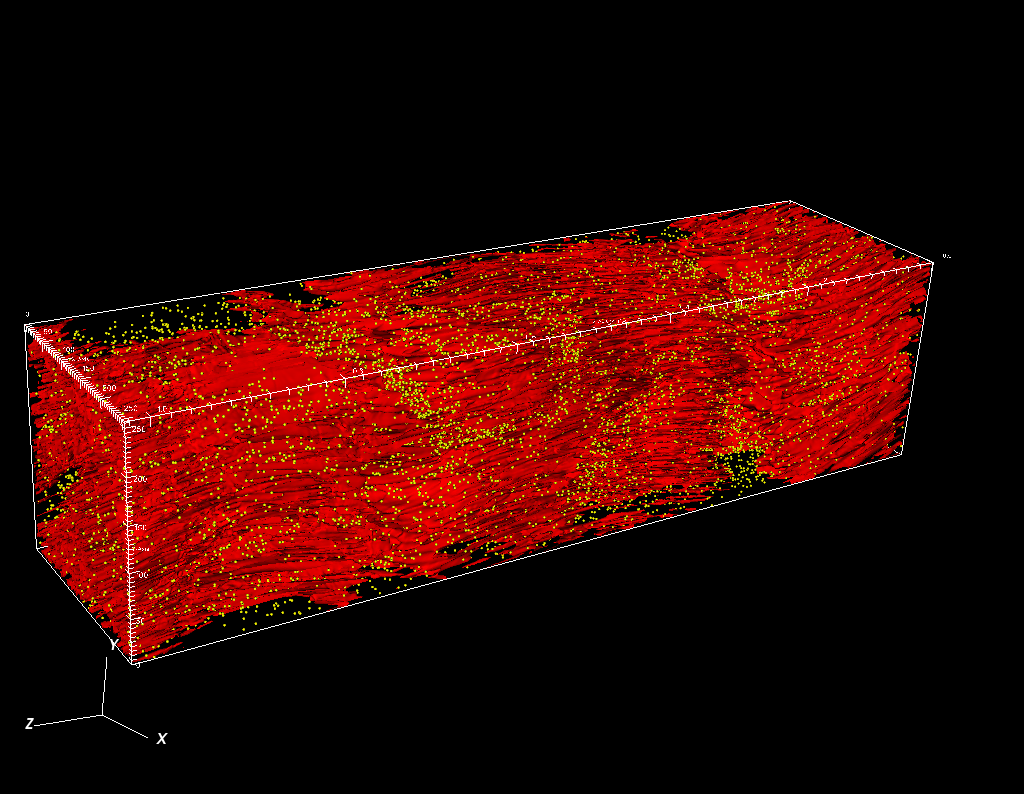}
\caption{Isosurface plots of the magnitude of the normal-fluid vorticity at
	$|\omega_n| $, for counterflow ST for a square cuboid
	simulation domain with $256 \times 256 \times 1024$ collocation points,
	$\widehat{\bf U}_{ns} = \hat{e_k}$, and $T=2.10$K. We indicate by small yellow 
	spheres the positions of particles (with $\rm St_n = 1.0 $).}
\label{fig:asym_box}
\end{figure}

\twocolumngrid

\end{document}